\documentclass[journal]{IEEEtran}
\IEEEoverridecommandlockouts

\usepackage[dvipsnames]{xcolor}
\usepackage{amsmath,amssymb,amsfonts}
\usepackage[linesnumbered,ruled]{algorithm2e}
\usepackage{algpseudocode}
\usepackage{graphicx}
\usepackage{textcomp}
\usepackage{xcolor}
\usepackage[inline]{enumitem}

\usepackage{multirow, multicol}
\usepackage{physics}    
\usepackage[flushleft]{threeparttable}
\usepackage{afterpage}
\usepackage[nolist]{acronym}

\usepackage{comment}
\usepackage[sort]{cite}
\usepackage[super]{nth}
\usepackage{multirow}       
\usepackage{amssymb}        
\usepackage{soul}           

\begin{acronym}
\acro{AI}{artificial intelligence}
\acro{IC}{integrated circuit}
\acro{DAS}{Dynamic Audio Sensor}
\acro{KWS}{keyword spotting}
\acro{VAD}{voice activity detection}
\acro{FEx}{feature extractor}
\acro{DFT}{discrete Fourier transform}
\acro{FFT}{fast Fourier transform}
\acro{DFT}{discrete Fourier transform}
\acro{AGC}{automatic gain control}
\acro{AFE}{analog front-end}
\acro{ADC}{analog-to-digital converter}
\acro{SNN}{spiking neural network}
\acro{FPGA}{field-programmable gate array}
\acro{ASIC}{application-specific integrated circuit}
\acro{SoC}{system-on-chip}
\acro{IoT}{Internet of Things}
\acro{LUT}{look-up table}
\acro{BPF}{band-pass filter}
\acro{LPF}{low-pass filter}
\acro{HPF}{high-pass filter}
\acro{OTA}{operational transconductance amplifier}
\acro{SF}{source-follower}
\acro{XSF}{cross-coupled source-follower}
\acro{SSF}{super source-follower}
\acro{FVF}{flipped voltage follower}
\acro{CCIA}{capacitively-coupled instrumentation amplifier}
\acro{DNN}{deep neural network}
\acro{RNN}{recurrent neural network}
\acro{MLP}{multilayer perceptron}
\acro{CT}{continuous-time}
\acro{DT}{discrete-time}
\acro{PWM}{pulse-width modulation}
\acro{PFM}{pulse-frequency modulation}
\acro{TDC}{time-to-digital converter}
\acro{PLL}{phase-locked loop}
\acro{ToF}{time-of-flight}
\acro{SAR}{successive approximation register}
\acro{SC}{switched-capacitor}
\acro{ECG}{electrocardiogram}
\acro{EEG}{electroencephalogram}
\acro{FIR}{finite impulse response}
\acro{ENOB}{effective number of bits}
\acro{RTL}{register-transfer level}
\acro{GSCD}{Google Speech Command Dataset}
\end{acronym}
\usepackage[draft]{pdfcomment}
\hypersetup{hidelinks}      

\begin{document}
\title{Continuous-Time Analog Filters for Audio Edge Intelligence:
Review on Circuit Designs}
\author{
Kwantae Kim, \IEEEmembership{Member, IEEE} and
Shih-Chii Liu, \IEEEmembership{Fellow, IEEE}
\thanks{This work was partially supported by the Swiss National Science Foundation BRIDGE project VIPS (181010) \emph{(Corresponding author: Shih-Chii Liu).}}
\thanks{The authors are with the Institute of Neuroinformatics, University of Zürich and ETH Zürich, 8057 Zürich, Switzerland (e-mail: kwantae@ini.uzh.ch, shih@ini.uzh.ch).}
\thanks{This paper is accepted for publication in IEEE Circuits and Systems Magazine. \copyright 2023 IEEE. Personal use of this material is permitted. Permission from IEEE must be obtained for all other uses, in any current or future media, including reprinting/republishing this material for advertising or promotional purposes, creating new collective works, for resale or redistribution to servers or lists, or reuse of any copyrighted component of this work in other works.}
}
\IEEEaftertitletext{\vspace{-2.0\baselineskip}}

\maketitle

\begin{abstract}
Edge audio devices can reduce data bandwidth requirements by pre-processing input speech on the device before transmission to the cloud. As edge devices are required to ensure \textit{always-on} operation, their stringent power constraints pose several design challenges and force IC designers to look for solutions that use low standby power. One promising bio-inspired approach is to combine the continuous-time analog filter channels with a small memory footprint deep neural network that is trained on edge tasks such as keyword spotting, thereby allowing all blocks to be embedded in an IC. This paper reviews the historical background of the continuous-time analog filter circuits that have been used as feature extractors for current edge audio devices. Starting from the interpretation of a basic biquad filter as a two-integrator-loop topology, we introduce the progression in the design of second-order low-pass and band-pass filters ranging from OTA-based to source-follower-based architectures. We also derive and analyze the small-signal transfer function and discuss their usage in edge audio applications.
\end{abstract}
\begin{IEEEkeywords}
Auditory, silicon cochlea, two-integrator-loop, band-pass filter (BPF), continuous-time (CT) filter, flipped voltage follower (FVF), second-order filter, source-follower (SF), super source-follower (SSF).
\end{IEEEkeywords}

\section{Introduction}

\begin{figure}[t]
    \begin{center}
    \includegraphics[width=\columnwidth]{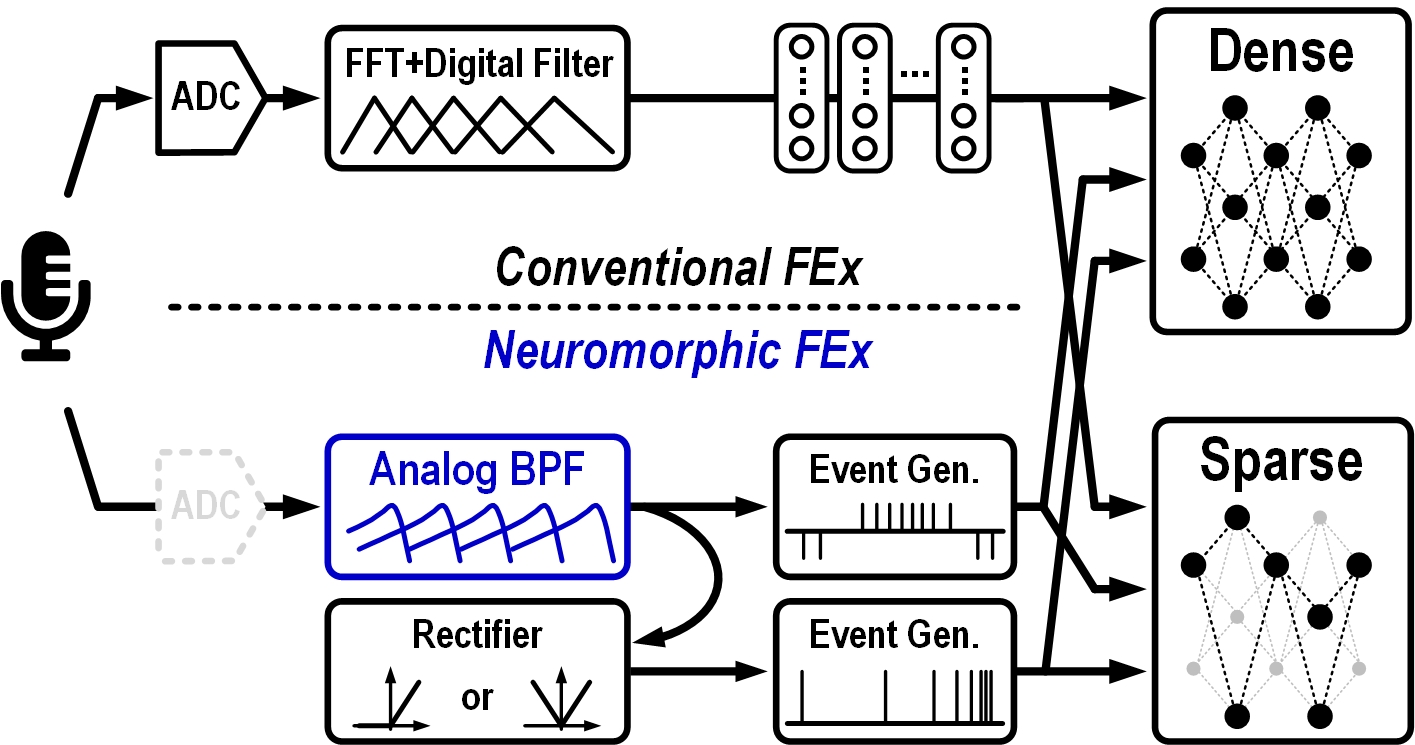}
    \caption{Edge audio processing stages based on conventional and neuromorphic approaches \cite{liu2021sscs_webinar, liu2022IEDM}.}\label{fig:block}
    \end{center}
    \vspace{-5mm}
\end{figure}

Edge audio devices are quickly gaining interest in the \ac{IoT} domain, with particular focus on low-power devices that perform smart pre-processing of the input before data transmission to the cloud. Typical tasks performed on these devices include \ac{VAD} and \ac{KWS}. {As shown in Fig.\,\,\ref{fig:block},} solutions for reported state-of-the-art edge audio integrated circuits (ICs) come in two forms. The first approach samples and quantizes the microphone output signal at Nyquist or oversampling frequency through an \ac{ADC}. These data samples are then further processed by a digital signal processing block such as \ac{FFT}, followed by triangular filtering, and logarithmic compression.

The second approach is to replace the synchronous \ac{ADC} and the subsequent signal processing stages with \ac{CT} analog circuits inspired by the biological modeling of cochleas \cite{lyon1988cochlea}. These designs implement the frequency-selective filtering properties of the basilar membrane, rectifying properties of the biological inner hair cells, and neuronal firing of the ganglion cells \cite{lyon2010history,Fragniere2005Cochlea,watts1992cochlea,liu2014cochlea,Wen2009cochlea,hamilton20082Dcochlea,Chan2007cochlea,abdalla2005ultrasonic}. Since the \ac{FFT} computation circuit is typically the most power-hungry building block of the entire audio \ac{FEx} \cite{giraldo2019kws}, the analog signal processing has been regarded as a promising alternative in terms of better power efficiency \cite{Sarpeshkar1998AvsD, Hall2005ProgAnalog} thereby it could be useful for tasks implemented on low-power edge audio devices. The \ac{CT} analog filters on the state-of-art edge audio ICs for \ac{VAD} \cite{Badami2016VAD, Yang2019VAD, Yang2021VAD} and \ac{KWS} \cite{Kim2022KWS, Wang2021KWS} adopt a set of second-order \acp{BPF}.

\begin{figure*}[t]
    \begin{center}
    \includegraphics[width=\textwidth]{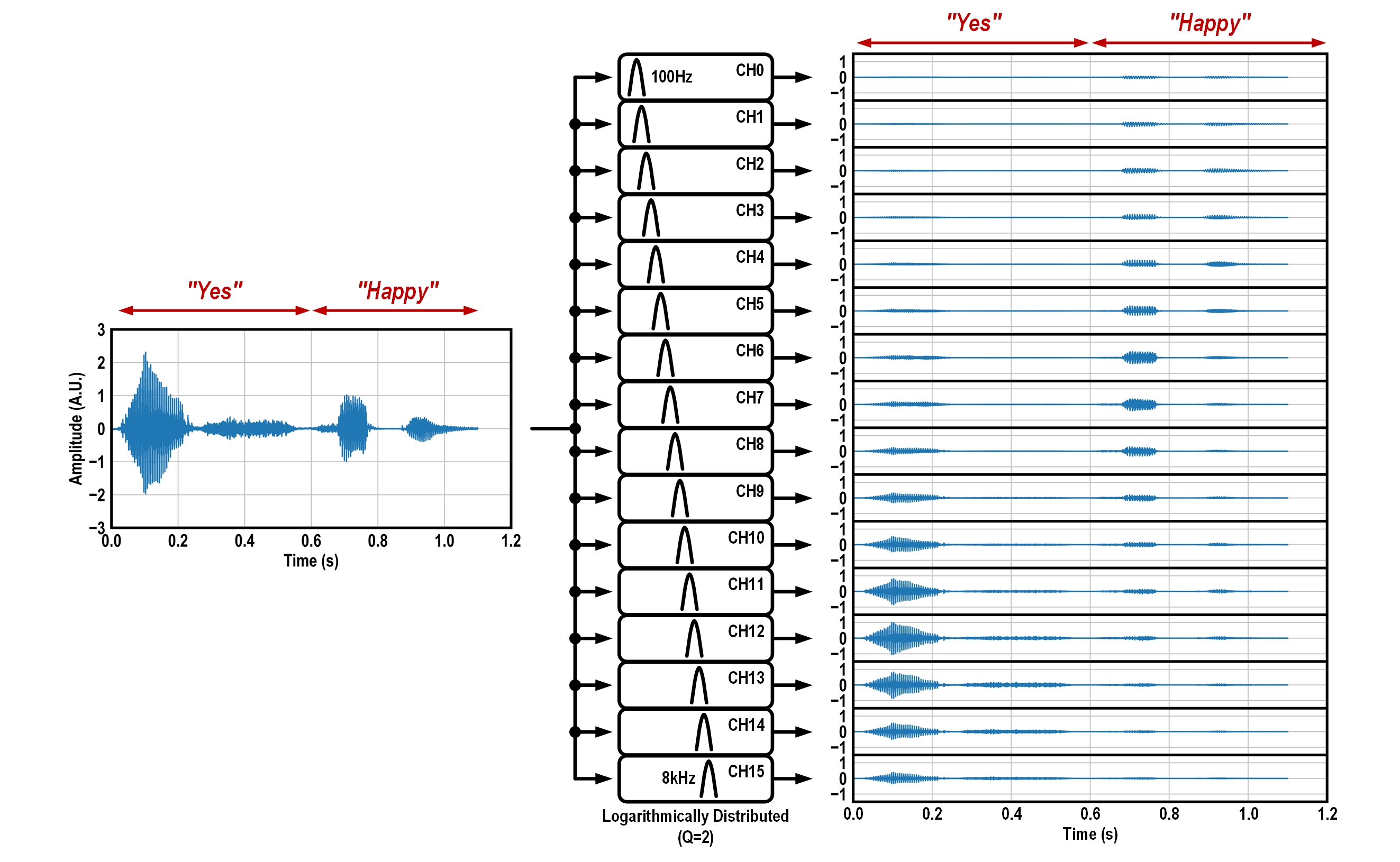}
    \caption{Example of the output of 16 bandpass filter bank channels with center frequencies ranging from 100 Hz to 8KHz on a log-spacing and with $Q=2$. The audio input is an example speech from the \ac{GSCD} samples.}\label{fig:GSCD}
    \end{center}
    \vspace{-2mm}
\end{figure*}

{Fig.\,\,\ref{fig:GSCD} shows an example of a \ac{CT} audio processing stage. Here, a speech sample from the \ac{GSCD} \cite{warden2018gscd} is fed to a 16-channel second-order Butterworth \ac{BPF} bank \cite{Kim2022KWS-JSSC}. It can be clearly seen that each channel responds to different parts of the speech sample depending on the instantaneous frequencies in the speech. These filter responses can be used for training a network on an audio task (see Fig\,\,\ref{fig:block}).}


This paper aims to provide an introductory survey of voltage-domain \ac{CT} analog filters leading to the circuits that have been reported in recent edge audio ICs. It will provide a unified analysis that covers {$g_\text{m}{C}$} and small-signal equivalent diagrams, based on a two-integrator-loop biquad topology. To the best of our knowledge, it is the first work to present the operating principle of voltage-domain second-order filters using an unified analysis that includes the \ac{OTA}-based, \ac{XSF}, \ac{SSF}, and \ac{FVF} biquad filters. Simulation results are also provided to show support for the proposed analysis. Note that the scope of this paper is geared to review transfer functions and thereby share intuitive circuit insights, rather than discussing every performance aspect of analog filter designs (e.g., noise, distortion, or sensitivity).


The remainder of this paper is organized as follows. Section\,\,\ref{sec:filter_review} introduces the basics of biquad filters and discusses how a second-order \ac{BPF} can be implemented from the two-integrator-loop topology. Section\,\,\ref{sec:declaration} presents the notation of a transconductor which is used in the description of the filters. Section\,\,\ref{sec:OTA_filter} and Section\,\,\ref{sec:SF_filter} present the core analysis of the \ac{OTA}-based and \ac{SF}-based filter circuits. Section\,\,\ref{sec:Summary} summarizes and discusses the state-of-the-art approaches for the design of edge audio ICs with brief future research prospects. Section\,\,\ref{sec:Conclusion} concludes this paper.

\begin{figure}[t]
    \begin{center}
    \includegraphics[width=0.9\columnwidth]{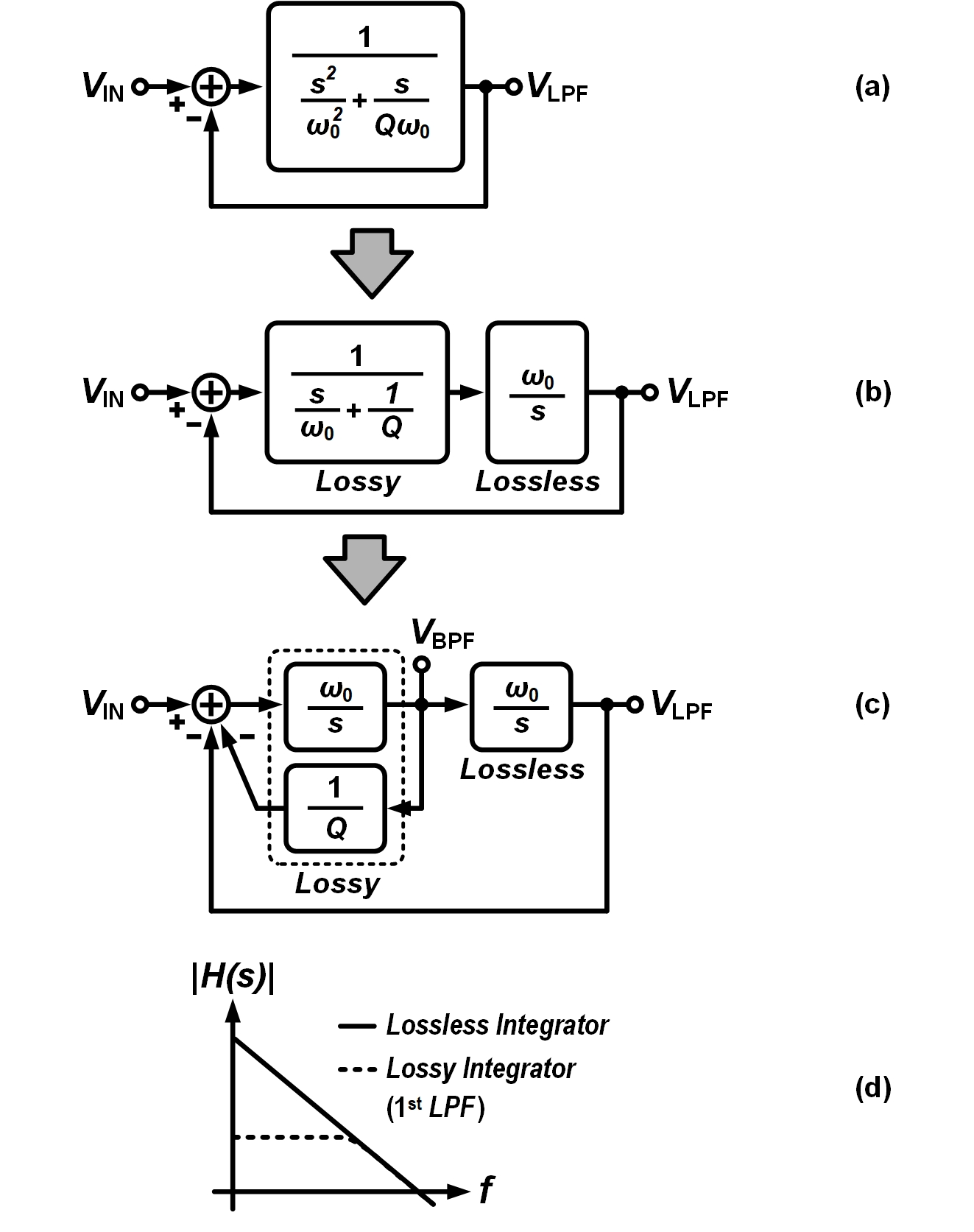}
    \caption{Two-integrator-loop representations of second-order \ac{LPF} and \ac{BPF}.}\label{fig:biquad}
    \end{center}
    \vspace{-3mm}
\end{figure}

\section{Review on Biquad Filter}\label{sec:filter_review}

The second-order filter is also called as a biquadratic filter or a biquad filter. It is because its transfer function is the ratio of two quadratic equations.

\paragraph{Biquad LPF}The generic transfer function of a second-order \ac{LPF} is
\begin{equation} \label{eq:biquad_LPF}
    H_\text{LPF}(s)=\frac{\displaystyle \omega_{0}^{2}}
    {\displaystyle s^{2}+\frac{\omega_{0}}{Q}s+\omega_{0}^{2}}
    =\frac{1}{\displaystyle \frac{s^{2}}{\omega_{0}^{2}}+\frac{s}{Q\omega_{0}}+1}
\end{equation}
where $\omega_{0}$ is the natural frequency (also the cutoff frequency in \ac{LPF} or center frequency in \ac{BPF}) and $Q$ is the quality factor of the filter. 
We demonstrate how this transfer function can be decomposed into multiple forms that lead to different circuit topologies. We first express Eq.\,\,(\ref{eq:biquad_LPF}) in a similar form to the transfer function of the closed-loop gain of a negative feedback system, i.e., $A_\text{CL}=A/(1+\beta A)$ where $A$ is the feedforward gain, $\beta$ is the feedback gain.
By setting $H_\text{LPF}(s)=A_\text{CL}(s)$, and defining $A(s)$ and $\beta(s)$ as
\begin{equation}
    A(s)=\frac{\displaystyle 1}{\displaystyle \displaystyle \frac{s^{2}}{\omega_{0}^{2}}+\frac{s}{Q\omega_{0}}}\qquad
    \beta (s)=1
\end{equation}
we can construct a corresponding block diagram as shown in Fig.\,\,\ref{fig:biquad}(a) where $H_\text{LPF}(s)=V_\text{LPF}(s)/V_\text{IN}(s)$.
This topology can be further decomposed into a cascaded structure forming a \emph{Lossy} integrator and a \emph{Lossless} integrator as shown in Fig.\,\,\ref{fig:biquad}(b) and is also called a \emph{Two-Integrator-Loop} topology \cite{sanchez-sinencio1988biquad}.
The characteristics of both integrator types are shown in Fig.\,\,\ref{fig:biquad}(d). We see that the first-order \ac{LPF} corresponds to the lossy case and an ideal integrator to the lossless case.
The lossy integrator can be further decomposed into a lossless integrator associated with a nested feedback path which controls $Q$ of the second-order filter as shown in Fig.\,\,\ref{fig:biquad}(c). This two-integrator-loop topology is used in a popular filter implementation called the Tow-Thomas biquad \cite{thomas1971biquad}.

\paragraph{Biquad BPF with Poles at the Same Frequency} Fig.\,\,\ref{fig:biquad}(c) also shows how a \ac{BPF} response is obtained at the output of the lossy integrator within this topology. This is because the lossless integrator is excluded in the feedforward gain of $H_\text{BPF}(s)$ compared to $H_\text{LPF}(s)$ case, which in turn acts as a differentiation of $H_\text{LPF}(s)$ (i.e., $\times s/\omega_{0}$). The transfer function of the resulting second-order \ac{BPF} is expressed as
\begin{equation} \label{eq:biquad_BPF}
    H_\text{BPF}(s)
    =\frac{\omega_{0}s}
    {\displaystyle s^{2}+\frac{\omega_{0}}{Q}s+\omega_{0}^{2}}
\end{equation}


\begin{figure}[t]
    \begin{center}
    \includegraphics[width=0.9\columnwidth]{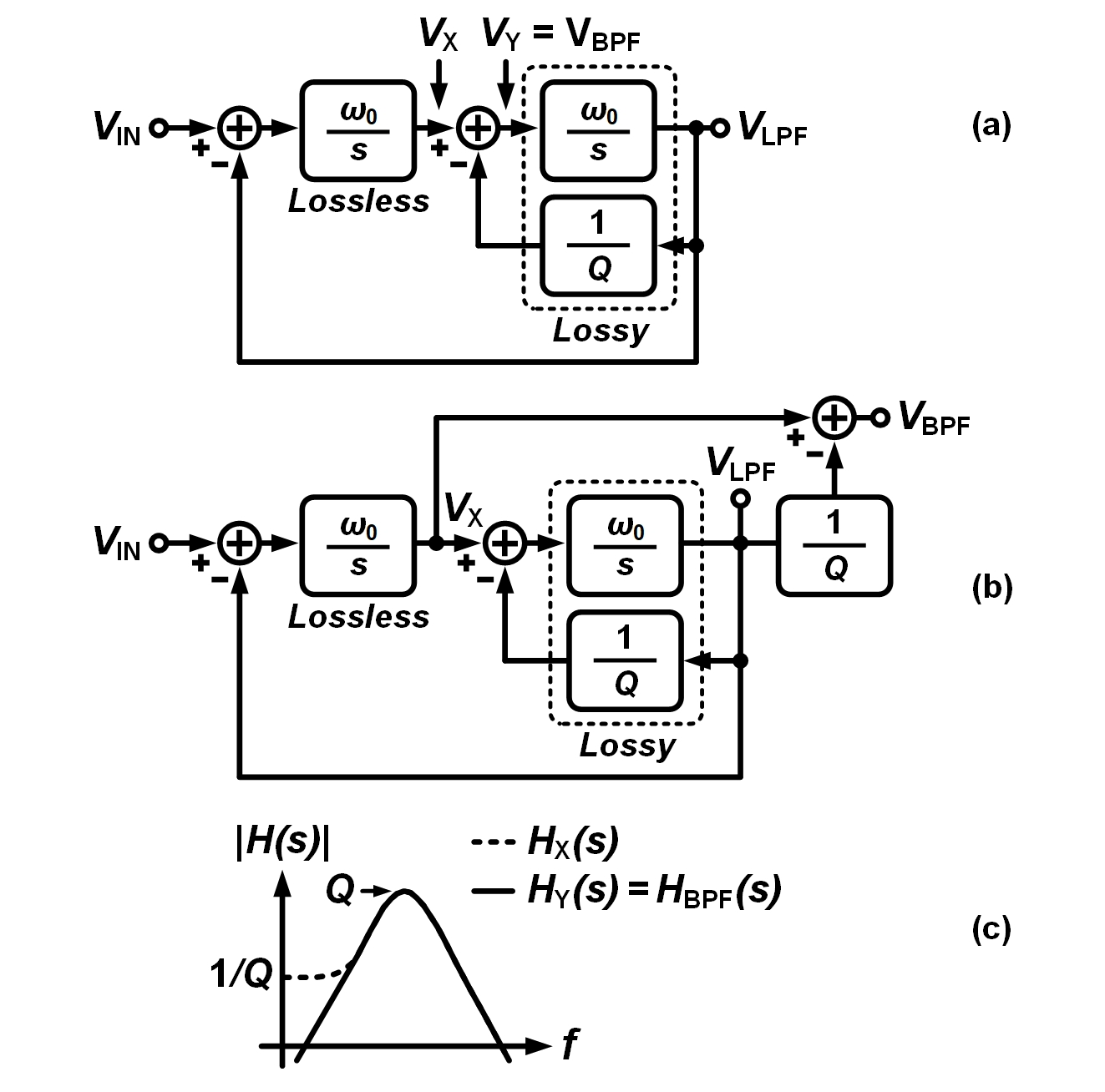}
    \caption{(a) Lossless-first two-integrator-loop, (b) design example of a \ac{BPF}, and (c) frequency response of $H_\text{X}(s), H_\text{Y}(s), H_\text{BPF}(s)$.}\label{fig:biquad_lossless-lossy1}
    \end{center}
    \vspace{-5mm}
\end{figure}

Fig.\,\,\ref{fig:biquad_lossless-lossy1}(a) shows a two-integrator-loop topology that uses a different configuration where the positions of the lossy and lossless integrators are reversed in contrast to the topology in Fig.\,\,\ref{fig:biquad}(c). It is clear that the transfer function $H_\text{LPF}(s)$ remains the same as in Eq.\,\,(\ref{eq:biquad_LPF}). 
This topology provides two additional output nodes, $V_\text{X}$ and $V_\text{Y}$. We first focus on the transfer function $H_\text{X}(s)$ 
because the subtraction operation which leads to $H_\text{Y}(s)$ typically happens within the differential-input transconductors in the continuous-time filters 
and thus $V_\text{Y}(s)$ cannot be extracted as an output. For example, $V_\text{Y}$ corresponds to $V_\text{GS}$ of the input transistor within the \ac{SF}-based filters (see Section\,\,\ref{subsec:sf}). Exceptions are type-I \ac{SSF} and type-I \ac{FVF} filters, which will be discussed in Sections\,\,\ref{subsec:SSF} and \ref{subsec:FVF}. The transfer function $H_\text{X}(s)$, which is extracted from $V_\text{X}$, the output of lossless integrator, deviates from the ideal \ac{BPF} ($H_\text{BPF}(s)$ in Eq.\,\,(\ref{eq:biquad_BPF})) response as an equation below.
\begin{equation} \label{eq:biquad_BPF_x}
    H_\text{X}(s)
    =\frac{\displaystyle \omega_{0}\left( s+\frac{\omega_{0}}{Q}\right)}
    {\displaystyle s^{2}+\frac{\omega_{0}}{Q}s+\omega_{0}^{2}}
\end{equation}

\begin{figure}[t]
    \begin{center}
    \includegraphics[width=0.9\columnwidth]{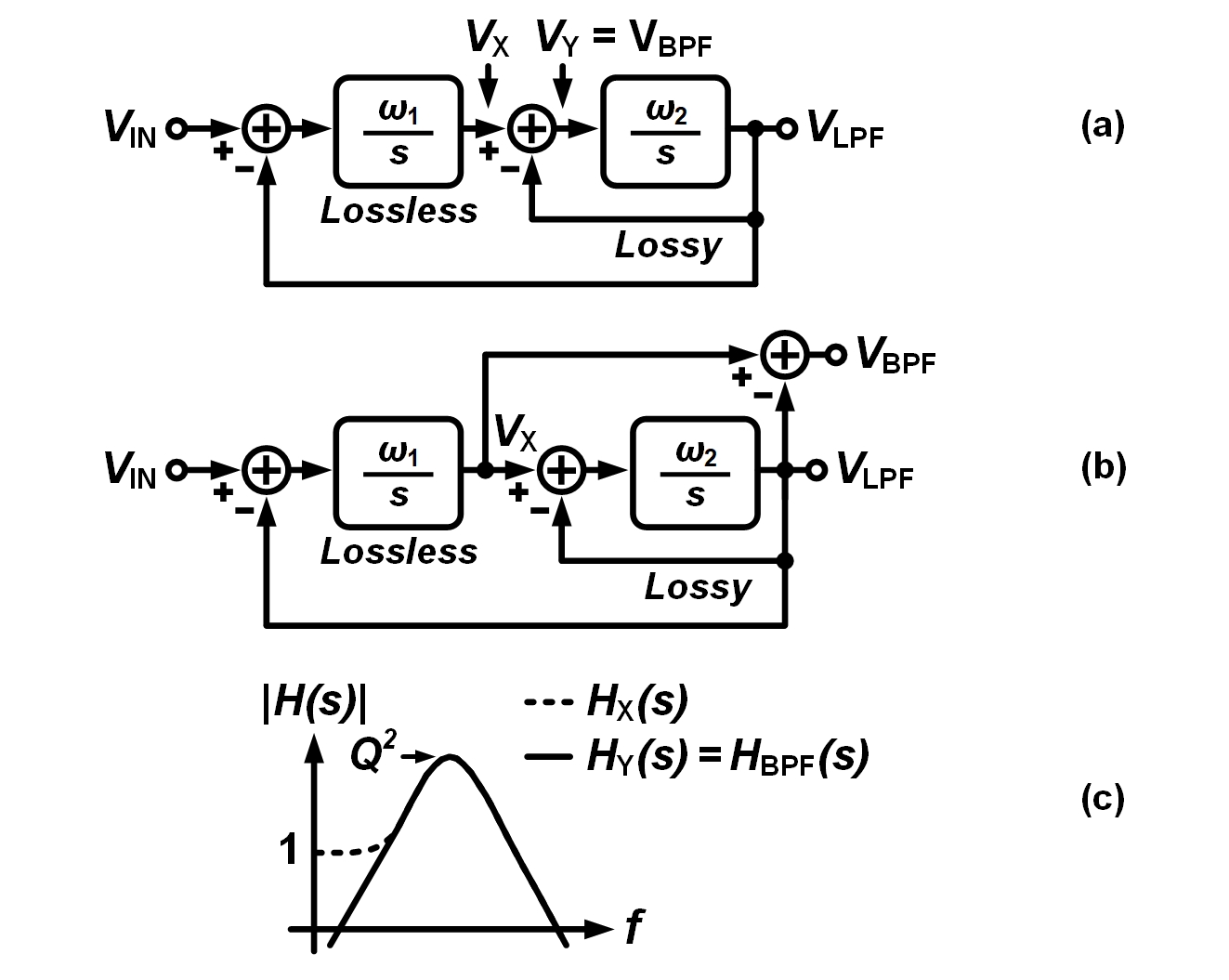}
    \caption{Two-integrator-loop representations with a lossless-first configuration using two different poles.}\label{fig:biquad_lossless-lossy2}
    \end{center}
    \vspace{-3mm}
\end{figure}

\begin{figure}[t]
    \begin{center}
    \includegraphics[width=0.9\columnwidth]{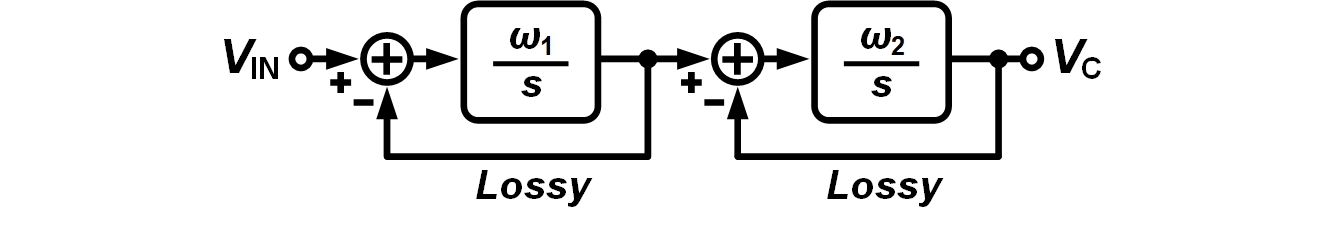}
    \caption{Second-order \ac{LPF} composed of cascaded lossy integrators.}\label{fig:lossy_cascade}
    \end{center}
    \vspace{-5mm}
\end{figure}

Fig.\,\,\ref{fig:biquad_lossless-lossy1}(c) shows that $s+\omega_{0}/Q$ term incurs lossy high-pass response in $H_\text{X}(s)$. This lossy response can also be predicted by calculating $H_\text{X}(0)$, which gives $1/Q$, in contrast to the case where $H_\text{BPF}(0)=0$. To achieve an output with an ideal \ac{BPF} ($H_\text{BPF}(s)$ in Eq.\,\,(\ref{eq:biquad_BPF})) curve using the lossless-first two-integrator-loop filter, we can use $V_\text{Y}$ (see Fig.\,\,\ref{fig:biquad_lossless-lossy1}(a)) where 
\begin{equation} \label{eq:biquad_BPF_y}
    H_\text{Y}(s)=H_\text{X}(s)-H_\text{LPF}(s)\frac{1}{Q}
    =H_\text{BPF}(s)
\end{equation}

Another method of getting a \ac{BPF} output using the lossless-first two-integrator-loop topology is shown in Fig.\,\,\ref{fig:biquad_lossless-lossy1}(b). Here, $V_\text{X}$ and $V_\text{LPF}$ are subtracted, outside the feedback loop, to obtain $V_\text{BPF}(s)$ as shown in (\ref{eq:biquad_BPF_y}).

\paragraph{Biquad BPF with Poles at Different Frequencies} We look next at the case where the filter has two different pole values (set by $\omega_{1}, \omega_{2}$) as in Fig.\,\,\ref{fig:biquad_lossless-lossy2} instead of two identical poles (Fig.\,\,\ref{fig:biquad_lossless-lossy1}). The transfer functions corresponding to the $V_\text{LPF}, V_\text{X}$ nodes and the $Q$ factor are described below.
\begin{equation}\label{eq:biquad_BPF_x_w1w2}\begin{split}
    H_\text{LPF}(s)&=\frac{\omega_{1}\omega_{2}}
    {s^{2}+\omega_{2}s+\omega_{1}\omega_{2}}\\
    H_\text{X}(s)&=\frac{\omega_{1}(s+\omega_{2})}
    {s^{2}+\omega_{2}s+\omega_{1}\omega_{2}}\\
    Q&=\sqrt{\frac{\omega_{1}}{\omega_{2}}}
\end{split}\end{equation}

Similarly to (\ref{eq:biquad_BPF_x}), $H_\text{X}(s)$ includes $s+\omega_{2}$ term in the numerator and thus it exhibits a lossy behavior at its high-pass shape as shown in Fig.\,\,\ref{fig:biquad_lossless-lossy2}(c) where $H_\text{X}(0)=1$ and its peak gain is $\omega_{1}/\omega_{2}=Q^{2}$. As in (\ref{eq:biquad_BPF_y}), by subtracting $V_\text{X}$ and $V_\text{LPF}$, one gets an ideal band-pass response.
\begin{equation}\label{eq:biquad_BPF_y_w1w2}\begin{split}
    H_\text{Y}(s)&=H_\text{X}(s)-H_\text{LPF}(s)\\
    &=\frac{\omega_{1}s}
    {s^{2}+\omega_{2}s+\omega_{1}\omega_{2}}=H_\text{BPF}(s)
\end{split}\end{equation}

The subtraction can be implemented either within the feedback loop ($V_\text{Y}$ in Fig.\,\,\ref{fig:biquad_lossless-lossy2}(a)) or out of the loop ($V_\text{BPF}$ in Fig.\,\,\ref{fig:biquad_lossless-lossy2}(b)). Note that $1/Q$ term is not multiplied with $H_\text{LPF}(s)$ in (\ref{eq:biquad_BPF_y_w1w2}) because $Q$ is now dependent on $\omega_{1}$ and $\omega_{2}$ while the feedback path within the lossy integrator in Fig.\,\,\ref{fig:biquad_lossless-lossy2} has unity gain.

\begin{figure}[t]
    \begin{center}
    \includegraphics[width=0.9\columnwidth]{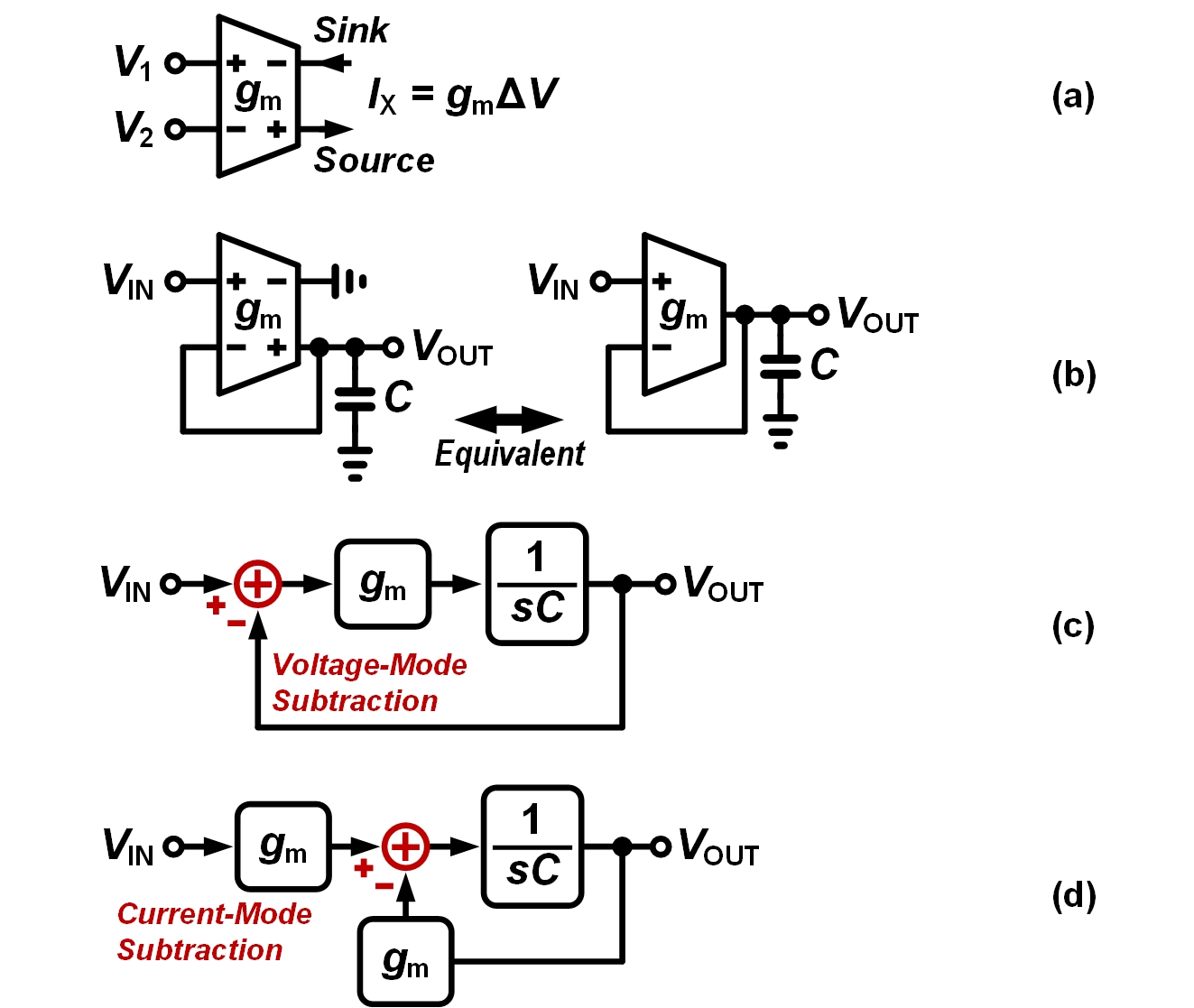}
    \caption{(a) The 4-port notation of a transconductor, (b) implementation of a lossy integrator with a transconductor, and block diagrams of a lossy integrator using (c) voltage-mode subtraction and (d) current-mode subtraction.}\label{fig:declare}
    \end{center}
    \vspace{-5mm}
\end{figure}

\begin{figure*}[t]
    \begin{center}
    \includegraphics[width=\textwidth]{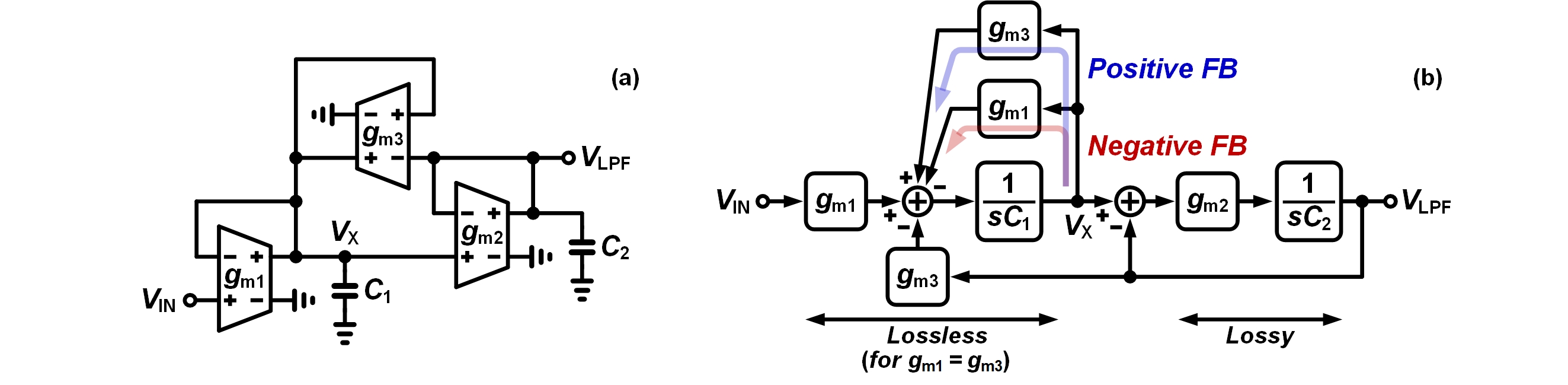}
    \caption{OTA-based second-order \ac{LPF} with (a) {$g_\text{m}{C}$} equivalent circuit and (b) small-signal diagram.}\label{fig:OTA-based-LPF}
    \end{center}
\end{figure*}


\paragraph{Biquad LPF with Cascaded Lossy Integrators} A second-order filter can alternatively be implemented by a cascade of two first-order lossy integrators with two different poles at ($\omega_{1}, \omega_{2}$) as shown in Fig.\,\,\ref{fig:lossy_cascade}. The transfer function of this topology can be calculated as below.
\begin{equation} \label{eq:lossy_cascade} \begin{split}
    H_\text{Cascade}(s)&=\frac{1}{\displaystyle 1+\frac{s}{\omega_{1}}}\cdot
    \frac{1}{\displaystyle 1+\frac{s}{\omega_{2}}}\\
    &=\frac{\omega_{1}\omega_{2}}{s^{2}+(\omega_{1}+\omega_{2})s+\omega_{1}\omega_{2}}\\
    Q &= \frac{\sqrt{\omega_{1}\omega_{2}}}{\omega_{1}+\omega_{2}}\leq 0.5
\end{split} \end{equation}

The $Q$-factor of this topology has a maximum value of 0.5, which one can {derive by using 
$x+y\geq 2\sqrt{xy}$}. In order to realize a wide $Q$ tunability, a second-order filter based on a two-integrator-loop topology is preferred over a cascaded lossy integrators.

In the following sections, we will analyze the second-order filter circuits by interpreting them either as lossy-first or as lossless-first two-integrator-loop topologies. In addition, we will only deal with the small-signal models excluding large-signal behaviors of the filter.

\section{Notation Declaration and Assumptions}\label{sec:declaration}

Throughout this paper, transconductors ($g_\text{m}$) will be depicted using a 4-port drawing as shown in Fig.\,\,\ref{fig:declare}(a) \cite{thanapitak2018buffer}, instead of the conventional 3-port drawing style that uses a single output port (rightside of Fig.\,\,\ref{fig:declare}(b)). This is particularly needed to analyze the source-follower-based filters, in which a single transistor acts as a transconductor and multiple transconductors are placed in a single bias current branch. We will discuss this type of filters in Section \ref{sec:SF_filter}. Fig.\,\,\ref{fig:declare}(b) shows an example of a {$g_\text{m}{C}$} lossy integrator implementation. The block diagram of this circuit can be described either with voltage-mode (Fig.\,\,\ref{fig:declare}(c)) subtraction or current-mode (Fig.\,\,\ref{fig:declare}(d)) subtraction. We will use both representations interchangeably.

{We assume that the intrinsic gain of the transistor is sufficiently large, i.e., $g_\text{m}r_\text{o}\gg 1$, where $r_\text{o}$ denotes the output impedance of a transistor, therefore, the load impedance of each transconductor can be approximated as $1/sC$. The body effect of the transistor is also ignored, i.e., $g_\text{mb}=0$.}

\section{OTA-Based Filters}\label{sec:OTA_filter}

\subsection{Second-Order LPF}\label{subsec:OTA_LPF}

Fig.\,\,\ref{fig:OTA-based-LPF} shows the \ac{OTA}-based second-order \ac{LPF} adopted in the early silicon cochlea designs \cite{mead1989a-vlsi, watts1992cochlea, lyon1988cochlea, Sarpeshkar1998Cochlea}.
The circuit consists of 3 \acp{OTA} and 2 capacitors leading to a {$g_\text{m}{C}$} topology. Note that the diode-connected source degeneration technique was used in the \acp{OTA} of the feed-forward path ($g_\text{m1}, g_\text{m2}$), to extend the linear input range of the filter \cite{watts1992cochlea} but at the expense of voltage headroom. The transfer function of this circuit is described {below.}
\begin{equation} \label{eq:OTA-based-LPF} \begin{split}
    H_\text{OTA-LPF}(s)=\frac{\displaystyle \frac{g_\text{m1}g_\text{m2}}{C_{1}C_{2}}}
    {\displaystyle s^{2}+s\left(
    \frac{g_\text{m1}}{C_{1}}
    -\frac{g_\text{m3}}{C_{1}}
    +\frac{g_\text{m2}}{C_{2}}
    \right)+\frac{g_\text{m1}g_\text{m2}}{C_{1}C_{2}}}\\
    \omega_{0}=\sqrt{\frac{g_\text{m1}g_\text{m2}}{C_{1}C_{2}}}\qquad
    Q=\frac{\displaystyle
    \sqrt{\frac{g_\text{m1}g_\text{m2}}{C_{1}C_{2}}}}
    {\displaystyle
    \frac{g_\text{m1}}{C_{1}}
    -\frac{g_\text{m3}}{C_{1}}
    +\frac{g_\text{m2}}{C_{2}}}
\end{split} \end{equation}

\begin{figure*}[t]
    \begin{center}
    \includegraphics[width=\textwidth]{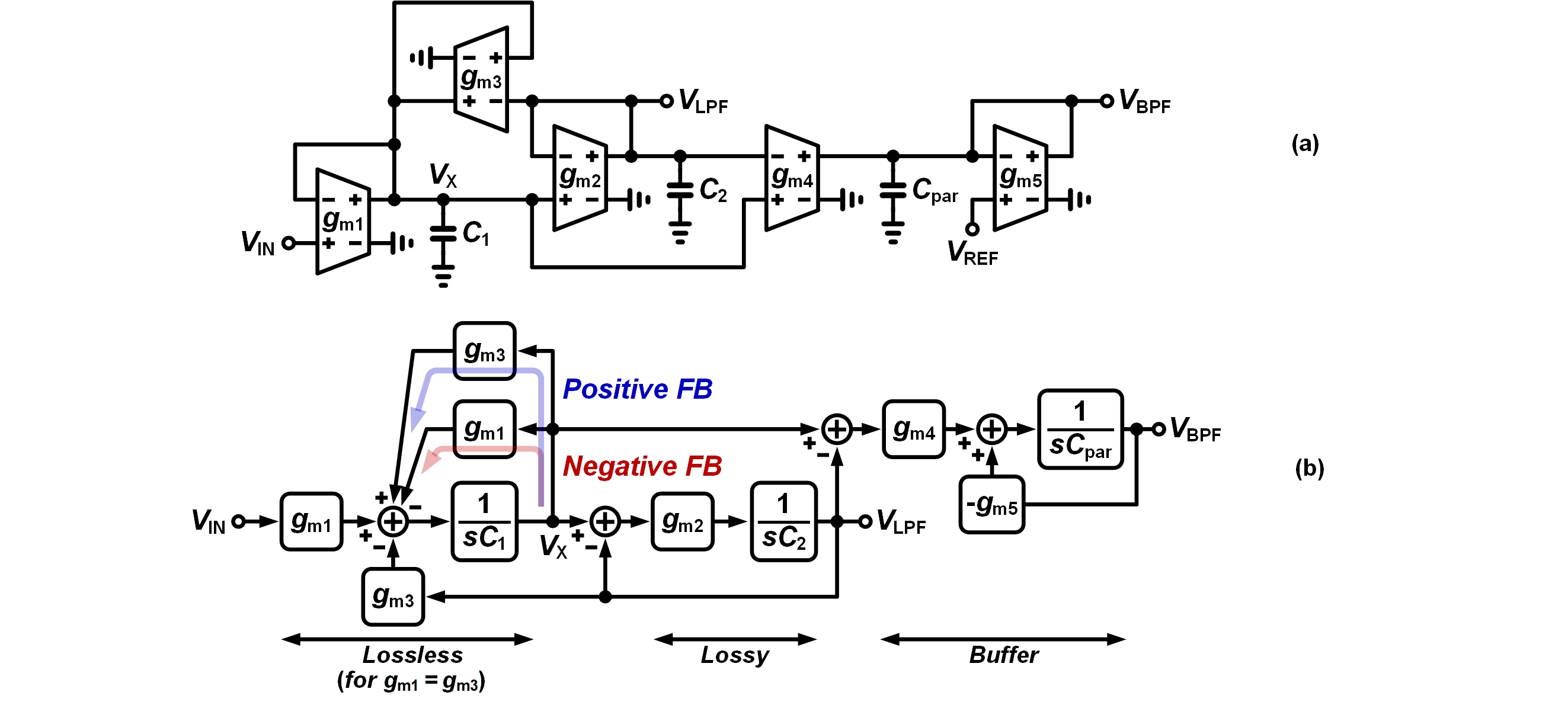}
    \caption{OTA-based second-order \ac{BPF} \cite{schaik1995cochlea} with (a) {$g_\text{m}{C}$} equivalent circuit and (b) small-signal diagram.}\label{fig:OTA-based-BPF}
    \end{center}
\end{figure*}


The basic structure of Fig.\,\,\ref{fig:OTA-based-LPF} is equivalent to the cascaded lossy integrators (or first-order \ac{LPF}s) in Fig.\,\,\ref{fig:lossy_cascade}. However, the added positive feedback $g_\text{m3}$ makes a significant difference to the transfer function in (\ref{eq:lossy_cascade}) because it cancels out the negative feedback within the first lossy integrator when $g_\text{m1}=g_\text{m3}$. It can be seen from the transfer function that $-g_\text{m3}/C_{1}$ cancels out $g_\text{m1}/C_{1}$ when $g_\text{m1}=g_\text{m3}$ thereby the overall topology reduces to the lossless-first two-integrator-loop structure (see Fig.\,\,\ref{fig:biquad_lossless-lossy2}(a)). In other words, the positive feedback converts a lossy integrator into a lossless integrator. This in turn leads to complex poles in its transfer function and its maximum $Q$ value is no longer limited to 0.5 in contrast to the case of cascaded lossy integrators (Fig.\,\,\ref{fig:lossy_cascade} and (\ref{eq:lossy_cascade})). The transfer function, $\omega_{0}$, and $Q$ factor of the \ac{OTA}-based \ac{LPF} 
when $g_\text{m1}=g_\text{m3}$ are given below.
\begin{equation} \label{eq:OTA-based_gm1gm3} \begin{split}
    H_\text{OTA-LPF1}(s)=\frac{\displaystyle \frac{g_{m1}g_{m2}}{C_{1}C_{2}}}
    {\displaystyle s^{2}+s\frac{g_{m2}}{C_{2}}+\frac{g_{m1}g_{m2}}{C_{1}C_{2}}}\\
    \omega_{0}=\sqrt{\frac{g_{m1}g_{m2}}{C_{1}C_{2}}}\qquad
    Q=\sqrt{\frac{g_{m1}C_{2}}{g_{m2}C_{1}}}
\end{split} \end{equation}
With this parameter setting, the positive feedback path is removed and thus the feedback stability is easier to be ensured.

In the original paper \cite{lyon1988cochlea} that proposed the \ac{LPF} in Fig.\,\,\ref{fig:OTA-based-LPF} for the cochlea channel, the following choices were made: $C=C_{1}=C_{2}$ and $g_\text{m}=g_\text{m1}=g_\text{m2}$ leading to the following equations for the transfer function, $Q$, and $\omega_{0}$. {The derived equation in (\ref{eq:OTA-based_C1C2gm1gm2}) is the same as the one introduced in \cite{lyon1988cochlea}.}
\begin{equation} \label{eq:OTA-based_C1C2gm1gm2} \begin{split}
    H_\text{OTA-LPF2}(s)=\frac{\displaystyle \frac{g_\text{m}^{2}}{C^{2}}}
    {\displaystyle s^{2}+s\left(
    \frac{2g_\text{m}}{C}-\frac{g_\text{m3}}{C}\right)
    +\frac{g_\text{m}^{2}}{C^{2}}}\\
    \omega_{0}=\frac{g_\text{m}}{C}\qquad
    Q=\frac{1}
    {\displaystyle 2\left( 1-\frac{g_\text{m3}}{2g_\text{m}} \right)}
\end{split} \end{equation}

Compared to (\ref{eq:OTA-based_gm1gm3}), this approach obtains $\omega_{0}$ and $Q$ with simpler forms. In addition, it offers better $\omega_{0}$ matching since the same $g_{m}$ and $C$ are used for both the first and second lossy integrators. However, the positive feedback path is not removed and thus design parameters must be chosen carefully to ensure stability.

\begin{figure*}[t]
    \begin{center}
    \includegraphics[width=\textwidth]{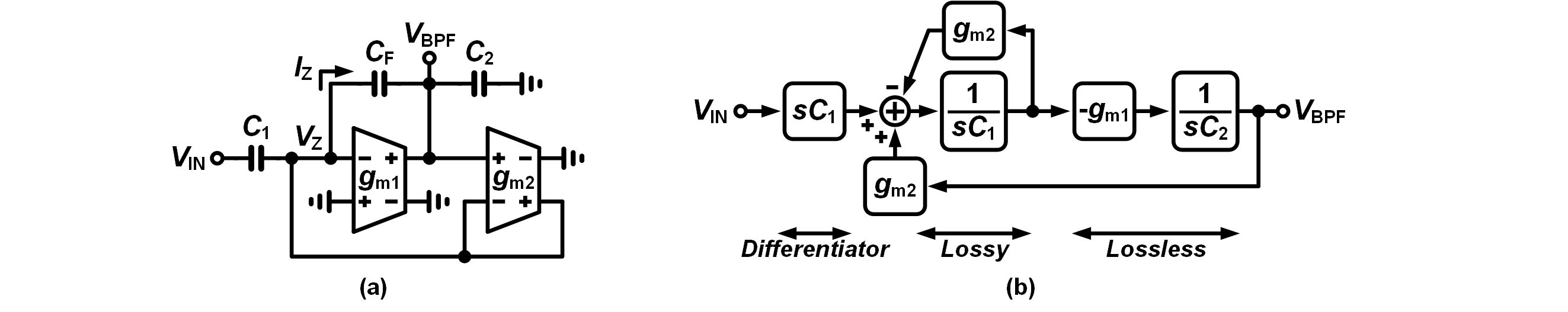}
    \caption{CCIA-based second-order \ac{BPF} \cite{Badami2016VAD} with (a) {$g_\text{m}{C}$} equivalent circuit and (b) small-signal diagram.}\label{fig:CCIA-based-BPF}
    \end{center}
\end{figure*}

\subsection{Second-Order BPF}\label{subsec:OTA_BPF}

Fig.\,\,\ref{fig:OTA-based-BPF} shows the implementation of an \ac{OTA}-based \ac{BPF} \cite{schaik1995cochlea}. In addition to the original \ac{LPF} in Fig.\,\,\ref{fig:OTA-based-LPF}, a differential buffer takes the difference of the $V_\text{X}$ and $V_\text{LPF}$. The buffer structure is equivalent to an open-loop first-order {$g_\text{m}{C}$} \ac{LPF} (see Chapter\,\,19 in \cite{sansen2006essential}) or a lossy integrator, however its cutoff frequency is set by a parasitic capacitance $C_\text{par}$. Note that the buffer is an open-loop design because the used transconductances for the feedforward ($g_\text{m4}$) and feedback ($g_\text{m5}$) paths are different.
This architecture follows the lossless-first two-integrator-loop topology with an external subtractor as discussed in Fig.\,\,\ref{fig:biquad_lossless-lossy2}(b). The transfer function of this \ac{OTA}-based \ac{BPF} is given in (\ref{eq:OTA-based-BPF})
\begin{equation} \label{eq:OTA-based-BPF} \begin{split}
    H_\text{OTA-BPF}(s)=\frac{\displaystyle s\frac{g_\text{m1}}{C_{1}}}
    {\displaystyle s^{2}+s\left(
    \frac{g_\text{m1}}{C_{1}}
    -\frac{g_\text{m3}}{C_{1}}
    +\frac{g_\text{m2}}{C_{2}}
    \right)+\frac{g_\text{m1}g_\text{m2}}{C_{1}C_{2}}}\\
    \omega_{0}=\sqrt{\frac{g_\text{m1}g_\text{m2}}{C_{1}C_{2}}}\qquad
    Q=\frac{\displaystyle
    \sqrt{\frac{g_\text{m1}g_\text{m2}}{C_{1}C_{2}}}}
    {\displaystyle
    \frac{g_\text{m1}}{C_{1}}
    -\frac{g_\text{m3}}{C_{1}}
    +\frac{g_\text{m2}}{C_{2}}}
\end{split} \end{equation}

which can be derived using a nodal analysis {in (\ref{eq:OTA-based-BPF-x})} where $H_\text{LPF}(s)$ is given in (\ref{eq:OTA-based-LPF}).

\begin{equation} \label{eq:OTA-based-BPF-x} \begin{split}
    H_\text{X}(s)&=\frac{H_\text{LPF}(s)}{H_\text{Lossy}(s)}
    =H_\text{LPF}(s)\left( 1+s\frac{C_{2}}{g_\text{m2}} \right)\\
    H_\text{Buf}(s)&=\frac{g_\text{m4}}{\displaystyle g_\text{m5}+sC_\text{par}}
    \approx 1\\
    H_\text{OTA-BPF}(s)&=\left( H_\text{X}(s)-H_\text{LPF}(s) \right)H_\text{Buf}(s)\\
    &=s\frac{C_{2}}{g_\text{m2}}H_\text{LPF}(s)
\end{split} \end{equation}

Here, the approximation for $H_\text{Buf}(s)$ is valid when the parasitic pole $g_\text{m5}/C_\text{par}$ within the buffer stage is far higher than $\omega_{0}$ of the \ac{BPF} and satisfying $g_\text{m4}=g_\text{m5}$. This means that the buffer stage is assumed to have a sufficiently wide bandwidth. Note that $\omega_{0}$ and $Q$ in (\ref{eq:OTA-based-BPF}) are the same as in (\ref{eq:OTA-based-LPF}) because the polynomial equation on the denominator does not change. Adopting the principle of this \ac{OTA}-based \ac{BPF} structure, further improved versions of the \ac{BPF} \cite{Chan2007cochlea, liu2014cochlea} that output a current-domain signal (excluding $g_\text{m5}$ in Fig.\,\,\ref{fig:OTA-based-BPF}) were implemented in a cascaded filter array. In \cite{Gao2019TIDIGITS}, the fabricated filter circuit \cite{liu2014cochlea} was used as an audio \ac{FEx} whose output was fed into a \ac{FPGA}-based \ac{RNN} classifier for speech recognition task using the TIDIGITS dataset.


A simpler form of the \ac{BPF} as shown in Fig.\,\,\ref{fig:CCIA-based-BPF} uses the \ac{OTA} as a core building block \cite{Badami2016VAD},{\cite{Shah2017FPAA}.}
This circuit adopts a \ac{CCIA} \cite{Harrison2003CCIA} associated with a buffer-based DC-servo loop. Its transfer function can be derived as in (\ref{eq:CCIA-based-BPF})
\begin{equation} \label{eq:CCIA-based-BPF} \begin{split}
    H_\text{CCIA-BPF}(s)=\frac{\displaystyle -s\frac{g_\text{m1}}{C_{2}}}
    {\displaystyle s^{2}+s\frac{g_\text{m2}}{C_{1}}
    +\frac{g_\text{m1}g_\text{m2}}{C_{1}C_{2}}}\\
    \omega_{0}=\sqrt{\frac{g_\text{m1}g_\text{m2}}{C_{1}C_{2}}}\qquad
    Q=\sqrt{\frac{g_{m1}C_{1}}{g_{m2}C_{2}}}
\end{split} \end{equation}
using a nodal analysis {in (\ref{eq:CCIA-based-BPF-nodal})}.
\begin{equation} \label{eq:CCIA-based-BPF-nodal} \begin{split}
    i_\text{Z}&=(v_\text{Z}-v_\text{BPF})sC_\text{F}\\
    &=(v_\text{IN}-v_\text{Z})sC_\text{1}+g_\text{m2}(v_\text{BPF}-v_\text{Z})\\
    \rightarrow v_\text{Z}&=\frac
    {(g_\text{m2}+sC_\text{F})v_\text{BPF}+sC_\text{1}v_\text{IN}}
    {g_\text{m2}+s(C_\text{F}+C_\text{1})}\\
    -g_\text{m1}v_\text{Z}&=sC_\text{2}v_\text{BPF}+(v_\text{BPF}-v_\text{Z})sC_\text{F}\\
    \rightarrow v_\text{Z}&=\frac{s(C_\text{F}+C_\text{2})}
    {sC_\text{F}-g_\text{m1}}v_\text{BPF}
\end{split} \end{equation}

Here, we assume $C_\text{1},C_\text{2}\gg C_\text{F}$ as used in \cite{Harrison2003CCIA}. We also assume the frequency range of this \ac{BPF} is far lower than $g_\text{m1}/C_\text{F}$ such that the filter parameters are solely determined by $C_\text{1},C_\text{2}$ capacitors and $g_\text{m1},g_\text{m2}$ transconductors. From Fig.\,\,\ref{fig:CCIA-based-BPF}(b), we see that this \ac{BPF} circuit has a lossy-first two-integrator-loop topology shown in Fig.\,\,\ref{fig:biquad}(c), with an added input differentiation stage $sC_{1}$. Therefore, although the block diagram might look like a second-order \ac{LPF}, the transfer function shows a \ac{BPF} response because of the differentiator. A parallel filter array adopting the \ac{BPF} topology in Fig.\,\,\ref{fig:CCIA-based-BPF} was implemented with an on-chip mixed-signal decision tree classifier in \cite{Badami2016VAD}, and demonstrated a 2-class \ac{VAD} task.

\section{Source-Follower-Based Filters}\label{sec:SF_filter}

\subsection{Source-Follower (First-Order LPF)}\label{subsec:sf}

\begin{figure}[t]
    \begin{center}
    \includegraphics[width=\columnwidth]{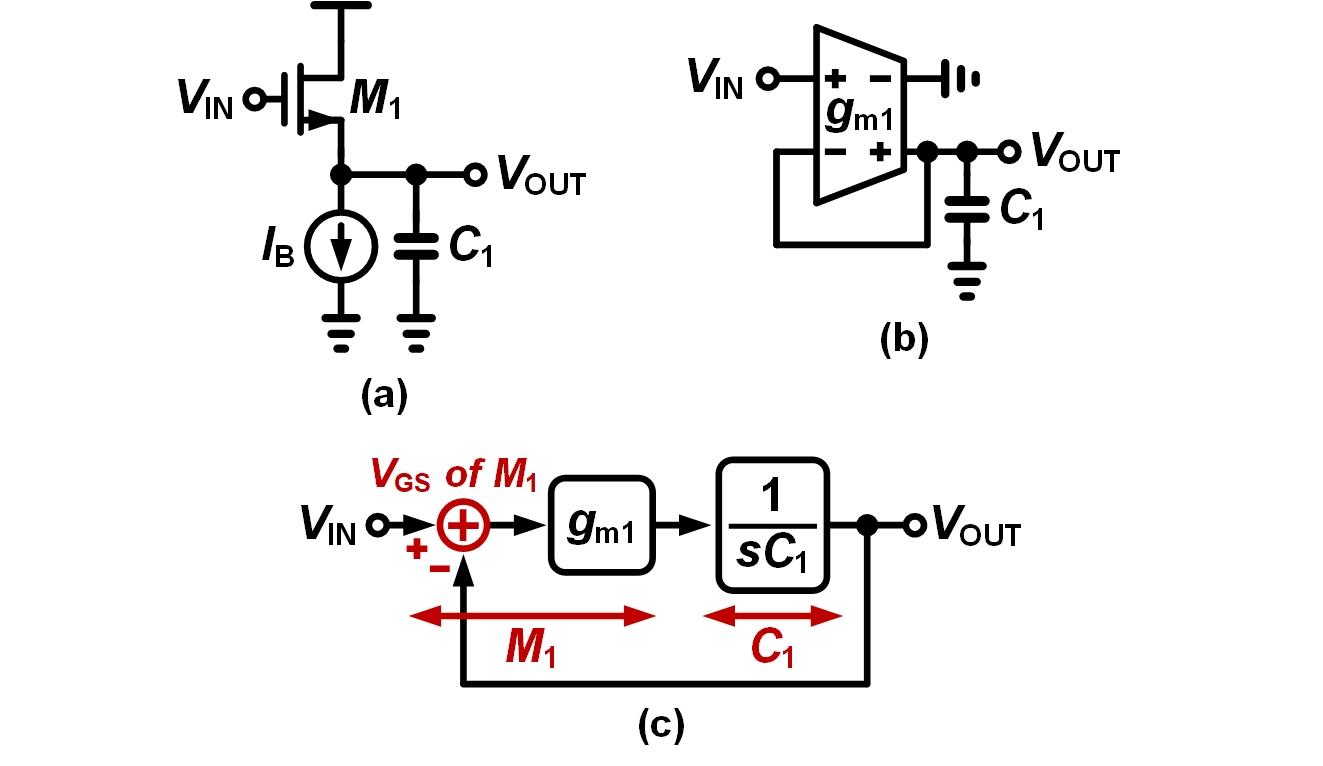}
    \caption{(a) Schematic, (b) {$g_\text{m}{C}$} equivalent circuit, and (c) small-signal diagram of the source-follower-based first-order \ac{LPF}.}\label{fig:sf-LPF}
    \end{center}
\end{figure}

Fig.\,\,\ref{fig:sf-LPF} shows a transistor-level schematic, {$g_\text{m}{C}$} equivalent circuit, and block diagram of the \ac{SF}-based first-order \ac{LPF}. The minus output port (sink) of $g_\text{m1}$ transconductor, which is the $M_{1}$ transistor, is tied to AC GND (corresponding to VDD in the schematic) in the {$g_\text{m}{C}$} equivalent circuit. The transfer function of the \ac{SF}-\ac{LPF} is given as below, denoting the equivalent transconductance of the circuit as $G_\text{m}=\partial I_\text{D}/\partial V_\text{IN}$ (see Chapter\,3.2.5 in \cite{razavi2001design}).
\begin{align}
    H_\text{SF}(s)&=G_\text{m}R_\text{L}=
    \frac{1}{\displaystyle \frac{1}{g_\text{m1}}+R_\text{L}
    \left(1+\frac{1}{g_\text{m1}r_\text{o1}}\right)}
    \cdot R_\text{L}\nonumber\\
    &=\frac{g_\text{m1}}{g_\text{m1}+g_\text{ds1}+g_\text{dsb}}\cdot
    \frac{1}{\displaystyle 1+s\frac{C_{1}}{g_\text{m1}+g_\text{ds1}+g_\text{dsb}}}
    \label{eq:sf-LPF_transfer}\\
    &\approx\frac{1}{\displaystyle 1+s\frac{C_\text{1}}{g_\text{m1}}}
    \label{eq:sf-LPF_transfer_final}
\end{align}

\begin{figure*}[t]
    \begin{center}
    \includegraphics[width=\textwidth]{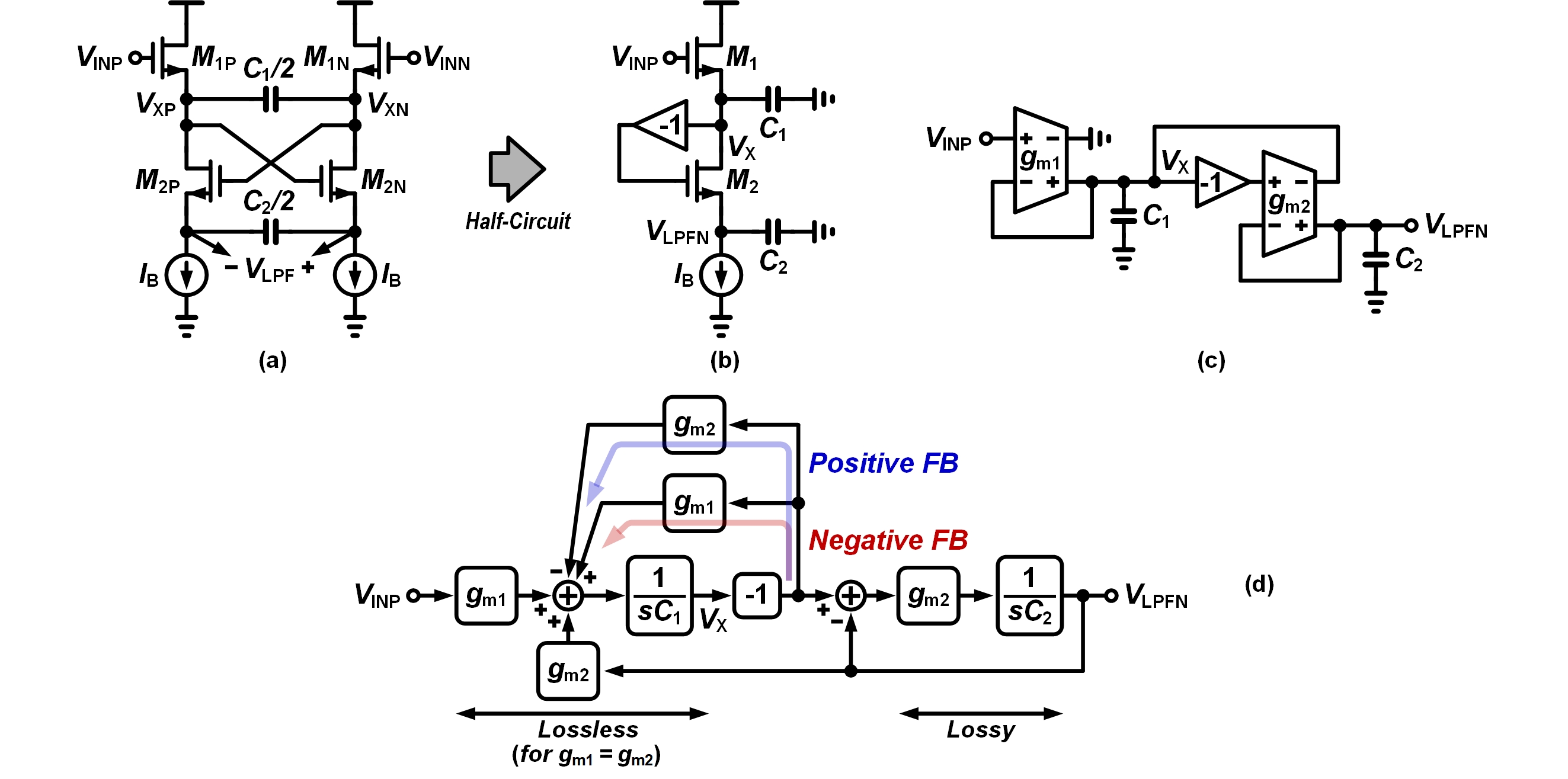}
    \caption{(a) Transistor-level schematic, (b) half-circuit representation of the schematic, (c) {$g_\text{m}{C}$} equivalent circuit, and (d) small-signal diagram of the cross-coupled source-follower-based \ac{LPF}.}\label{fig:XSF-LPF}
    \end{center}
    \vspace{-5mm}
\end{figure*}
\begin{figure*}[!t] \begin{equation} \label{eq:XSF-LPF} \begin{split}
    H_\text{XSF-LPF}(s)=\frac{\displaystyle \frac{g_\text{m1}g_\text{m2}}{C_{1}C_{2}}}
    {\displaystyle s^{2}+s\left(
    \frac{g_\text{m1}}{C_{1}}
    -\frac{g_\text{m2}}{C_{1}}
    +\frac{g_\text{m2}}{C_{2}}
    \right)+\frac{g_\text{m1}g_\text{m2}}{C_{1}C_{2}}}\qquad
    \omega_{0}=\sqrt{\frac{g_\text{m1}g_\text{m2}}{C_{1}C_{2}}}\qquad
    Q=\frac{\displaystyle
    \sqrt{\frac{g_\text{m1}g_\text{m2}}{C_{1}C_{2}}}}
    {\displaystyle
    \frac{g_\text{m1}}{C_{1}}
    -\frac{g_\text{m2}}{C_{1}}
    +\frac{g_\text{m2}}{C_{2}}}
\end{split} \end{equation} \end{figure*}

Here, $r_\text{o1}=1/g_\text{ds1}$ is the output impedance of transistor $M_{1}$, $r_\text{ob}=1/g_\text{dsb}$ is the output impedance of tail current source, and $R_\text{L}=(1/sC_\text{1})||r_\text{ob}=1/(sC_\text{1}+g_\text{dsb})$ is the load impedance seen at the $V_\text{OUT}$ node. The approximation in (\ref{eq:sf-LPF_transfer_final}) is valid assuming sufficiently large intrinsic gain of the transistor ($g_\text{m}r_\text{o}=g_\text{m}/g_\text{ds}\gg 1$). Since its cutoff frequency $\omega_\text{0}=g_\text{m1}/C_\text{1}$ includes transconductance and capacitance, the \ac{SF}-\ac{LPF} is a first-order g\textsubscript{m}C filter or a lossy integrator as discussed in Section \ref{sec:filter_review} and presented in Fig.\,\,\ref{fig:declare}(b). As shown in Fig.\,\,\ref{fig:sf-LPF}(b), the \ac{SF} implements a local feedback around the source node of the $M_\text{1}$ transistor. Because the closed-loop gain at DC is given in the first term of (\ref{eq:sf-LPF_transfer}), we can derive a DC loop gain $T_\text{0}$ of the local feedback in the \ac{SF}, using the equation of closed-loop gain in the negative feedback system $A_\text{CL}=A/(1+\beta A)$ where $T=\beta A$ and $\beta=1$ are used considering the unity-gain nature of the \ac{SF}.
\begin{equation} \label{eq:sf-LPF_loopgain} \begin{split}
    A_\text{CL0}&=\frac{g_\text{m1}}{g_\text{m1}+g_\text{ds1}+g_\text{dsb}}
    =\frac{\displaystyle \frac{g_\text{m1}}{g_\text{ds1}+g_\text{dsb}}}
    {\displaystyle 1+\frac{g_\text{m1}}{g_\text{ds1}+g_\text{dsb}}}\\
    T_\text{0}&=\frac{g_\text{m1}}{g_\text{ds1}+g_\text{dsb}}
    =g_\text{m1}\cdot r_\text{o1}||r_\text{ob}
\end{split} \end{equation}

The above equation shows that the closed-loop gain $A_\text{CL0}$ approaches 1 as $T_\text{0}\gg 1$. In other words, the linearity performance of a \ac{SF}-\ac{LPF} is enhanced with a larger transconductance $g_\text{m1}$ benefiting from its local feedback. Compared to the conventional open-loop g\textsubscript{m}C filters in which the input transistors are usually operated at strong-inversion to suppress harmonic distortions \cite{sansen2006essential}, the \ac{SF}-\ac{LPF} allows the input transistor to operate in subthreshold. Since the subthreshold region offers a higher $g_\text{m}$ within a given power budget, i.e., higher $g_\text{m}/I_\text{D}$, the \ac{SF}-based filters can be a promising option in terms of linearity-noise trade-offs. Note that the active RC filters also operate with a negative feedback loop thereby offering higher linearity than open-loop g\textsubscript{m}C filters, however they have a higher power burden because of the needed amplifiers to drive resistive loads \cite{sansen2006essential}. Overall, the \ac{SF} filters are more suitable for ultra-low-power audio filter implementation.

\subsection{Cross-Coupled Source-Follower (Second-Order LPF)}\label{subsec:CCSF-LPF}

Fig.\,\,\ref{fig:XSF-LPF} shows a transistor-level full schematic, half-circuit equivalent, {$g_\text{m}{C}$} equivalent, and block diagram of the \ac{XSF}-based second-order \ac{LPF} \cite{amico2006sf-LPF, Zhang2013XSF} while the transfer function and filter parameters are given in (\ref{eq:XSF-LPF}). Note that we define the polarity of the \ac{LPF} output in a reversed way due to the cross-coupled structure within the filter circuit. For instance, the signal flow starting from $V_\text{INP}$ ends at $V_\text{LPFP}$ through the source-following operations of $M_\text{1P}$ and $M_\text{2N}$. If we would like to extract the output from the left-half of the circuit ($V_\text{LPFN}$) while also keeping the input fixed to the left-half ($V_\text{INP}$), an ideal inverting buffer is required considering differential structure of the circuit, as shown in the half-circuit schematic Fig.\,\,\ref{fig:XSF-LPF}(b).

Although not easy to identify, interestingly, the core structure of this filter circuit is the same as the \ac{OTA}-based \ac{LPF} in Fig.\,\,\ref{fig:OTA-based-LPF}, i.e., it has a lossless-first two-integrator-loop topology with auxiliary positive and negative feedback paths. Assuming $g_\text{m1}=g_\text{m2}$ as used in the original paper \cite{amico2006sf-LPF}, the transfer function, $\omega_{0}$, and $Q$ of \ac{XSF} filter are given  below:
\begin{equation} \label{eq:XSF-LPF_transfer_gm1gm2} \begin{split}
    H_\text{XSF1}(s)=\frac
    {\displaystyle \frac{g_\text{m1}^{2}}{C_{1}C_{2}}}
    {\displaystyle s^{2}
    +s\frac{g_\text{m1}}{C_{2}}
    +\frac{g_\text{m1}^{2}}{C_{1}C_{2}}}\\
    \omega_{0}=\frac{g_\text{m1}}{\sqrt{C_{1}C_{2}}}\qquad
    Q=\sqrt{\frac{C_{2}}{C_{1}}}
\end{split} \end{equation}
This equation becomes quite similar to the transfer function of the \ac{OTA}-based \ac{LPF} assuming $g_\text{m1}=g_\text{m2}=g_\text{m3}$ in (\ref{eq:OTA-based_gm1gm3}).

\begin{figure}[t]
    \begin{center}
    \includegraphics[width=\columnwidth]{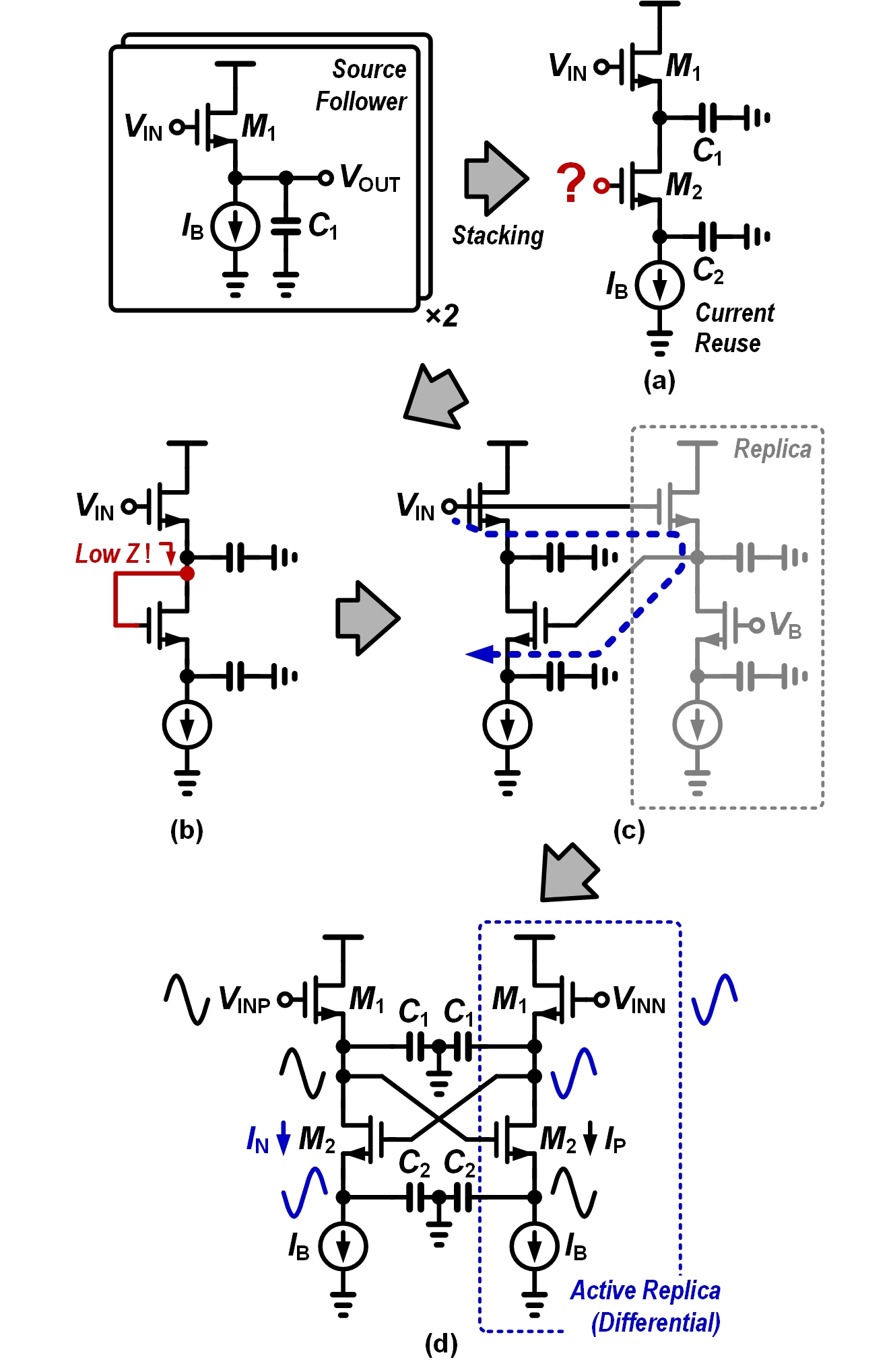}
    \caption{A step-by-step procedure for implementing the cross-coupled source-follower-based filter.}\label{fig:XSF-LPF-insight}
    \end{center}
    \vspace{-3mm}
\end{figure}

To provide an intuitive insight into the operation of the \ac{XSF} circuit, a step-by-step design procedure starting from a basic \ac{SF} is illustrated in Fig.\,\,\ref{fig:XSF-LPF-insight}. First, let us consider the case of stacking two \ac{SF}s in a single branch. The stacking is especially beneficial as it allows a better $g_\text{m}$ matching and easier cut-off frequency tuning, because the bias current of the two \acp{SF} are reused. However, if the two \acp{SF} are stacked, the input of the second \ac{SF} should be connected to the output of the first \ac{SF}, which in turn, requiring gate and drain ports of input transistor in the second \ac{SF} to be shorted into a single node. It results in a diode-connection of the input transistor as shown in Fig.\,\,\ref{fig:XSF-LPF-insight}(b), which incurs a low-impedance load to the first \ac{SF} output, thereby leading to a failure of proper source-following operation of the first \ac{SF}. One possible solution to this problem is using a replica \ac{SF} (Fig.\,\,\ref{fig:XSF-LPF-insight}(c)). Here, input of the main circuit goes to the input of the replica circuit and we assume that the gate of the $M_{2}$ transistor in replica circuit is properly biased ($V_\text{B}$). Note that the output of the first \ac{SF} now has a cascode current source which is a high-impedance load. Therefore, the output of the second \ac{SF} forms a cascaded lossy integrator (the blue line in Fig.\,\,\ref{fig:XSF-LPF-insight}(c)) as discussed in Fig.\,\,\ref{fig:lossy_cascade}. Finally, we can exploit this replica circuit as an active building block that operates in a complementary manner to the main circuit, i.e., differential circuit, using a cross-coupling technique. The final schematic of the \ac{XSF} is drawn as in Fig.\,\,\ref{fig:XSF-LPF-insight}(d). Note that the added positive feedback path (uppermost $g_\text{m2}$ in Fig.\,\,\ref{fig:XSF-LPF}(d)) cancels lossy property of the first \ac{SF} within the cascaded lossy integrator and thus the maximum $Q$ value is no longer limited to 0.5, as discussed in Section \ref{sec:OTA_filter}. We can omit the GND connection between the two $C_{1}$ capacitors and merge them into a single $C_{1}/2$ capacitor like in Fig.\,\,\ref{fig:XSF-LPF}(a) because the input signal is in differential-mode and thus a virtual GND forms when $C_{1}/2$ in Fig.\,\,\ref{fig:XSF-LPF}(a) is splitted into the two serialized $C_{1}$.

\begin{equation} \label{eq:XSF-LPF_transfer_x_gm1gm2} \begin{split}
    H_\text{X}(s)&=\frac{H_\text{XSF1}(s)}{H_\text{Lossy}(s)}
    =H_\text{XSF1}(s)\left( 1+s\frac{C_{2}}{g_\text{m1}} \right)\\ 
    &=\frac{\displaystyle \frac{g_\text{m1}^{2}}{C_{1}C_{2}}
    \left( 1+s\frac{C_{2}}{g_\text{m1}} \right)}
    {\displaystyle s^{2}
    +s\frac{g_\text{m1}}{C_{2}}
    +\frac{g_\text{m1}^{2}}{C_{1}C_{2}}}
\end{split} \end{equation}

The transfer function for $V_\text{X}$ node is derived as {in (\ref{eq:XSF-LPF_transfer_x_gm1gm2})}. As discussed in Fig.\,\,\ref{fig:biquad_lossless-lossy2}(a) and (\ref{eq:biquad_BPF_x_w1w2}), $H_\text{X}(s)$ exhibits a lossy behavior at its low-frequency band.

\begin{figure}[t]
    \begin{center}
    \includegraphics[width=\columnwidth]{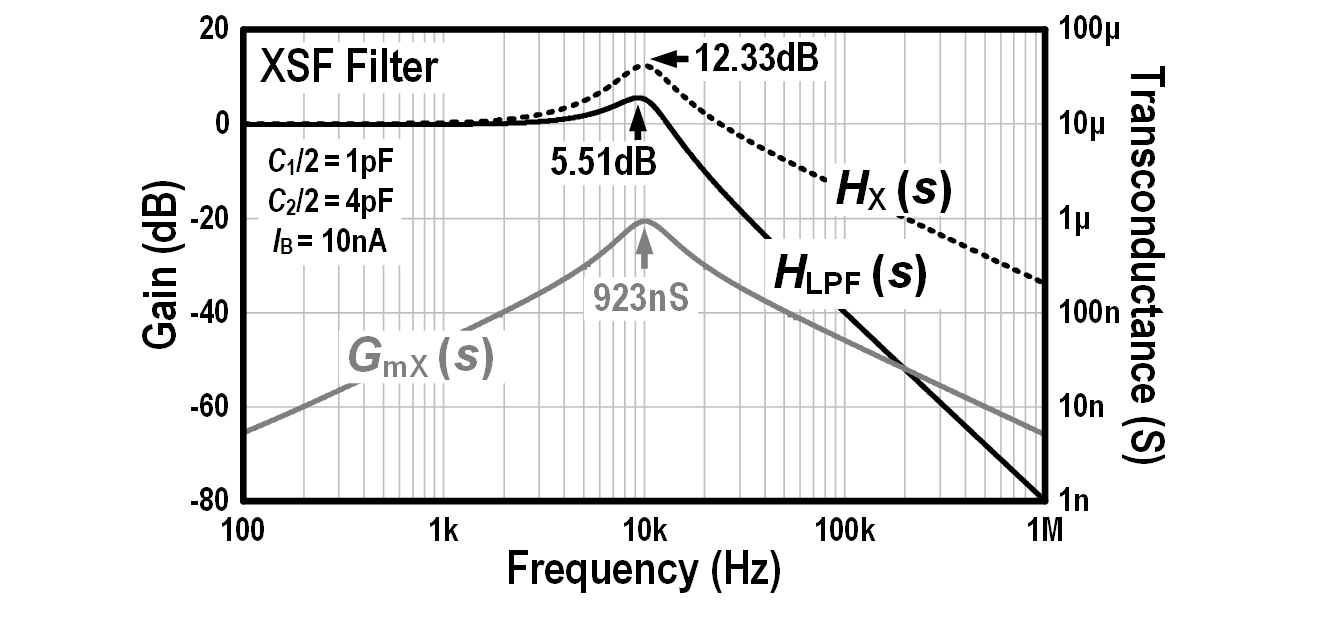}
    \caption{Simulated frequency response of the \ac{XSF} filter.}\label{fig:XSF-LPF-sim}
    \end{center}
    \vspace{-5mm}
\end{figure}

\begin{figure*}[t]
    \begin{center}
    \includegraphics[width=\textwidth]{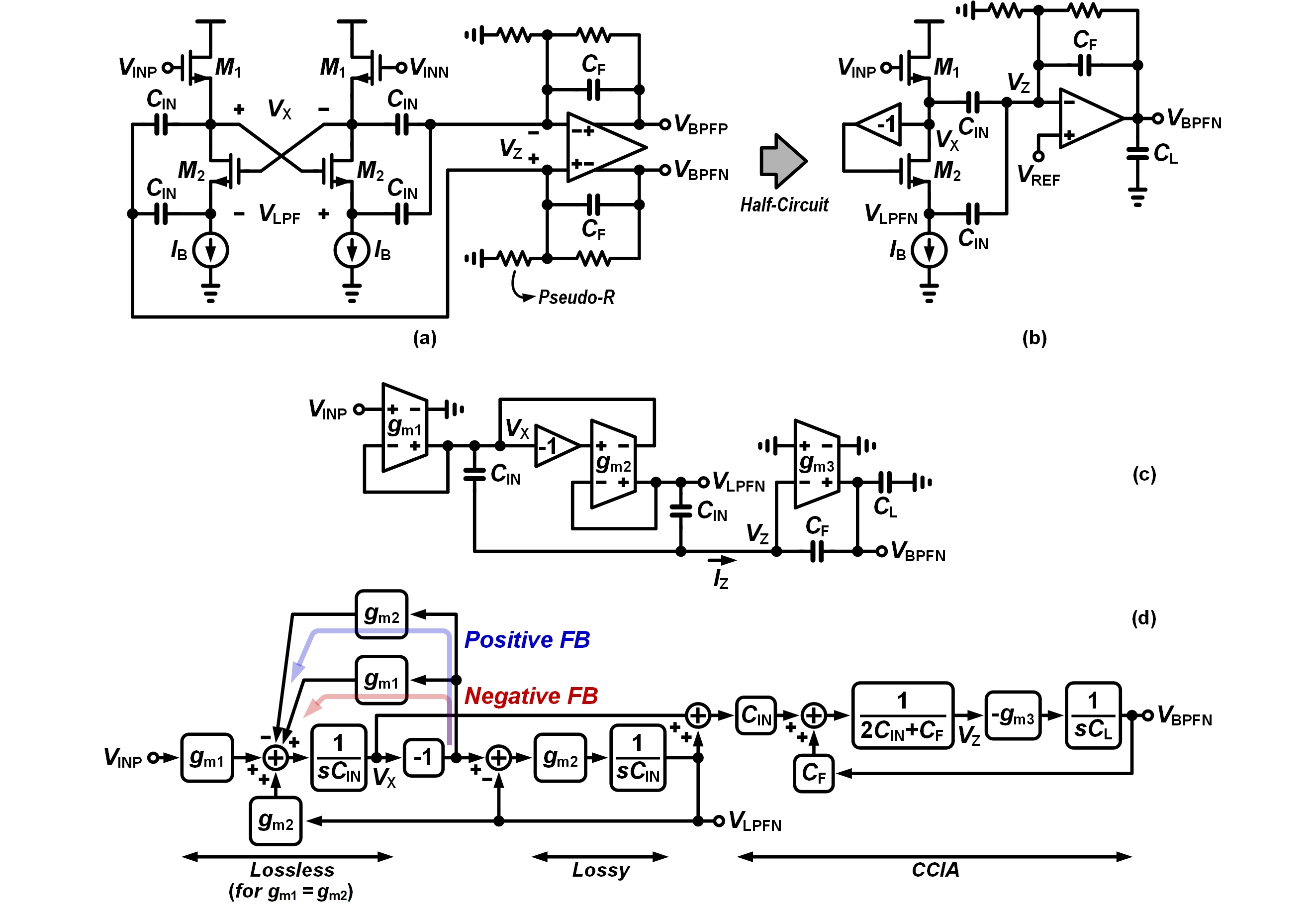}
    \caption{(a) Transistor-level schematic, (b) half-circuit representation of the schematic, (c) {$g_\text{m}{C}$} equivalent circuit, and (d) small-signal diagram of the cross-coupled source-follower-based \ac{BPF} \cite{Yang2016Cochlea}.}\label{fig:XSF-BPF}
    \end{center}
    \vspace{-5mm}
\end{figure*}
\begin{figure*}[!t] \begin{equation} \label{eq:XSF-BPF} \begin{split}
    H_\text{XSF-BPF}(s)=\frac{C_\text{IN}}{C_\text{F}}\cdot
    \frac{\displaystyle s\frac{g_\text{m1}}{C_\text{IN}}}
    {\displaystyle s^{2}+s\frac{g_\text{m1}}{C_\text{IN}}
    +\frac{g_\text{m1}g_\text{m2}}{C_\text{IN}^{2}}}\qquad
    \omega_{0}=\frac{\sqrt{g_\text{m1}g_\text{m2}}}{C_\text{IN}}\qquad
    Q=\sqrt{\frac{g_\text{m2}}{g_\text{m1}}}
\end{split} \end{equation} \end{figure*}

To evaluate the frequency response at $V_\text{X}$ and the analysis discussed in Fig.\,\,\ref{fig:biquad_lossless-lossy2} and (\ref{eq:biquad_BPF_x_w1w2}), an AC simulation is conducted with \ac{XSF} filter circuit using a 65-nm CMOS process. The width and length of all transistors are sized at 1\,$\mu m$.
The supply voltage is set as 1.2\,V. Also, high $V_\text{TH}$ devices are used since they have higher intrinsic gain than nominal $V_\text{TH}$ devices. The source and body contacts of the transistors are shorted to negate body effects (a deep N-well layout is required in this case), therefore $g_\text{m1}=g_\text{m2}$ and (\ref{eq:XSF-LPF_transfer_x_gm1gm2}) becomes valid. We will use the same simulation setups in Sections\,\,\ref{subsec:SSF} and \ref{subsec:FVF} unless otherwise it is specified. We set $C_{2}/2=4C_{1}/2=4\,\text{pF}$ such that $Q = 2$. Fig.\,\,\ref{fig:XSF-LPF-sim} shows frequency responses of $H_\text{LPF}(s), H_\text{X}(s), \ \text{and} \ G_\text{mX}(s)$ where
\begin{equation} \label{eq:gmx}
G_\text{mX}(s)=\frac{i_\text{P}-i_\text{N}}{v_\text{INP}-v_\text{INN}}(s)
\end{equation}
stands for the transconductance of differential current flowing through $M_{2}$ transistor in Fig.\,\,\ref{fig:XSF-LPF-insight}(d). As expected, $H_\text{LPF}(s)$ shows a second-order \ac{LPF} behavior with a 40dB/dec roll-off while $H_\text{X}(s)$ has a band-pass characteristic but having a lossy behavior on its low-frequency band. Since we have designed $Q=2$, the peak gain is expected as $Q^{2}=4=12.04\,\text{dB}$ according to Fig.\,\,\ref{fig:biquad_lossless-lossy2} which closely matches to $12.33\,\text{dB}$ in our simulation result. The plotted $G_\text{mX}(s)$ graph shows a \ac{BPF} response which corresponds to the output of $g_\text{m2}$ within the lossy integrator in Fig.\,\,\ref{fig:XSF-LPF}(d), as also discussed in (\ref{eq:biquad_BPF_y_w1w2}). This is because the $V_\text{Y}$ node in (\ref{eq:biquad_BPF_y_w1w2}) corresponds to the output of the second subtractor in Fig.\,\,\ref{fig:XSF-LPF}(d). As the extracted DC operating point in our simulation shows $g_\text{m2}=253.2\,\text{nS}$, the peak value of frequency response in $G_\text{mX}(s)$ is expected as $4\times253.2\,\text{nS}=1012.8\,\text{nS}$ where $Q^{2}=4$ is considered according to (\ref{eq:biquad_BPF_x_w1w2}, \ref{eq:biquad_BPF_y_w1w2}). This estimation also makes a close agreement with $923\,\text{nS}$ from our simulation result. The center frequency $f_{0}$ can be estimated using (\ref{eq:XSF-LPF}) as follow,
\begin{equation}
f_{0}=\frac{\omega_{0}}{2\pi}
=\frac{\sqrt{(253.2\,\text{nS})^{2}}}
{2\pi\sqrt{2\times 1\,\text{pF}\times 2\times 4\,\text{pF}}}
=10.07\,\text{kHz}
\end{equation}
which also makes an agreement with $9.98\,\text{kHz}$ from the simulation result in Fig.\,\,\ref{fig:XSF-LPF-sim}. The residual estimation errors may come from parasitic capacitance and insufficient $g_\text{m}r_\text{o}$ as discussed in (\ref{eq:sf-LPF_loopgain}).

\subsection{Cross-Coupled Source-Follower (Second-Order BPF)}\label{subsec:CCSF-BPF}

Similarly to Fig.\,\,\ref{fig:OTA-based-BPF} \cite{schaik1995cochlea}, a \ac{BPF} architecture with an external subtractor applied to a \ac{XSF} was proposed in \cite{Yang2016Cochlea} as shown in Fig.\,\,\ref{fig:XSF-BPF}. We use the same \ac{XSF} circuit from Fig.\,\,\ref{fig:XSF-LPF} to describe the operating principle for consistency, despite a folded input stage was used in \cite{Yang2016Cochlea}. The circuit uses a \ac{CCIA} \cite{Harrison2003CCIA} to subtract $V_\text{LPF}$ from $V_\text{X}$. The input ports of the \ac{OTA} ($V_\text{Z}$) within a \ac{CCIA} works as a virtual GND by the negative feedback. Therefore, the virtual GND nodes that also exist in the \ac{XSF} (see Fig.\,\,\ref{fig:XSF-LPF-insight}(d)) can be reused. In effect, $C_\text{IN}$ capacitors in Fig.\,\,\ref{fig:XSF-BPF}(a) contribute to the following: (1) filtering capacitors of the \ac{XSF}; (2) input capacitors of the \ac{CCIA}. The \ac{XSF} circuit uses $C_\text{1}=C_\text{2}=C_\text{IN}$ to realize $V_\text{X}-V_\text{LPF}$ equation within the \ac{CCIA}, otherwise it forms a weighted addition, i.e., $C_{1}V_\text{X}-C_{2}V_\text{LPF}$. We assume that the resistance of the pseudo-resistor is sufficiently large such that the associated AC-coupling high-pass cut-off frequency stays far smaller than $\omega_{0}$ in \ac{BPF}. This allows us to omit pseudo-resistors in a small-signal {$g_\text{m}{C}$} equivalent circuit shown in Fig.\,\,\ref{fig:XSF-BPF}(c). Based on the nodal analysis below, we can derive the transfer function of the \ac{CCIA} $H_\text{CCIA}(s)$.
\begin{align}
    i_\text{Z}&=[(v_\text{X}-v_\text{Z})+(-v_\text{LPF}-v_\text{Z})]sC_\text{IN}\nonumber\\
    &=[v_\text{Z}-(-v_\text{BPF})]sC_\text{F}\nonumber\\
    -g_\text{m3}v_\text{Z}&=-v_\text{BPF}sC_\text{L}
    +(-v_\text{BPF}-v_\text{Z})sC_\text{F}\nonumber\\
    H_\text{CCIA}(s)&=\frac{v_\text{BPF}}{v_\text{X}-v_\text{LPF}}(s)\nonumber\\
    &\approx\frac{C_\text{IN}}{C_\text{F}}\cdot
    \frac{1}{\displaystyle 1+s\frac{C_\text{L}}{g_\text{m3}\cdot
    \frac{\displaystyle C_\text{F}}{\displaystyle 2C_\text{IN}+C_\text{F}}}}
    \approx \frac{C_\text{IN}}{C_\text{F}}\label{eq:XSF-BPF-CCIA}
\end{align}
where $i_\text{Z}$ represents a small-signal current in Fig.\,\,\ref{fig:XSF-BPF}(c), $C_\text{L}$ is a load capacitance of the \ac{CCIA}. We assume $C_\text{L}\gg C_\text{F}$ and $s\ll g_\text{m3}/C_\text{F}$ for the first approximation in (\ref{eq:XSF-BPF-CCIA}). We can see that the bandwidth of \ac{CCIA} is reduced by the amount of feedback factor $\beta=C_\text{F}/(2C_\text{IN}+C_\text{F})$ from the original \ac{OTA} bandwidth $g_\text{m3}/C_\text{L}$, as also described in \cite{Harrison2003CCIA}. The second approximation used in (\ref{eq:XSF-BPF-CCIA}) is valid when the bandwidth of \ac{CCIA} is sufficiently higher than $\omega_{0}$ in the \ac{BPF}. The transfer function of the \ac{XSF}-based \ac{BPF} $H_\text{BPF}(s)$ is derived in (\ref{eq:XSF-BPF}) using the equations as below. $H_\text{LPF}(s)$ can be found in (\ref{eq:XSF-LPF}) but with an additional condition of $C_\text{1}=C_\text{2}=C_\text{IN}$.
\begin{equation}\label{eq:XSF-BPF-derive}\begin{split}
    H_\text{X}(s)&=\frac{H_\text{LPF}(s)}{H_\text{Lossy}(s)}
    =H_\text{LPF}(s)\left( 1+s\frac{C_\text{IN}}{g_\text{m2}} \right)\\
    H_\text{BPF}(s)&=\left( H_\text{X}(s)-H_\text{LPF}(s) \right)H_\text{CCIA}(s)\\
    &=s\frac{C_\text{IN}}{g_\text{m2}}H_\text{LPF}(s)\cdot
    \frac{C_\text{IN}}{C_\text{F}}
\end{split}\end{equation}

\begin{figure*}[t]
    \begin{center}
    \includegraphics[width=\textwidth]{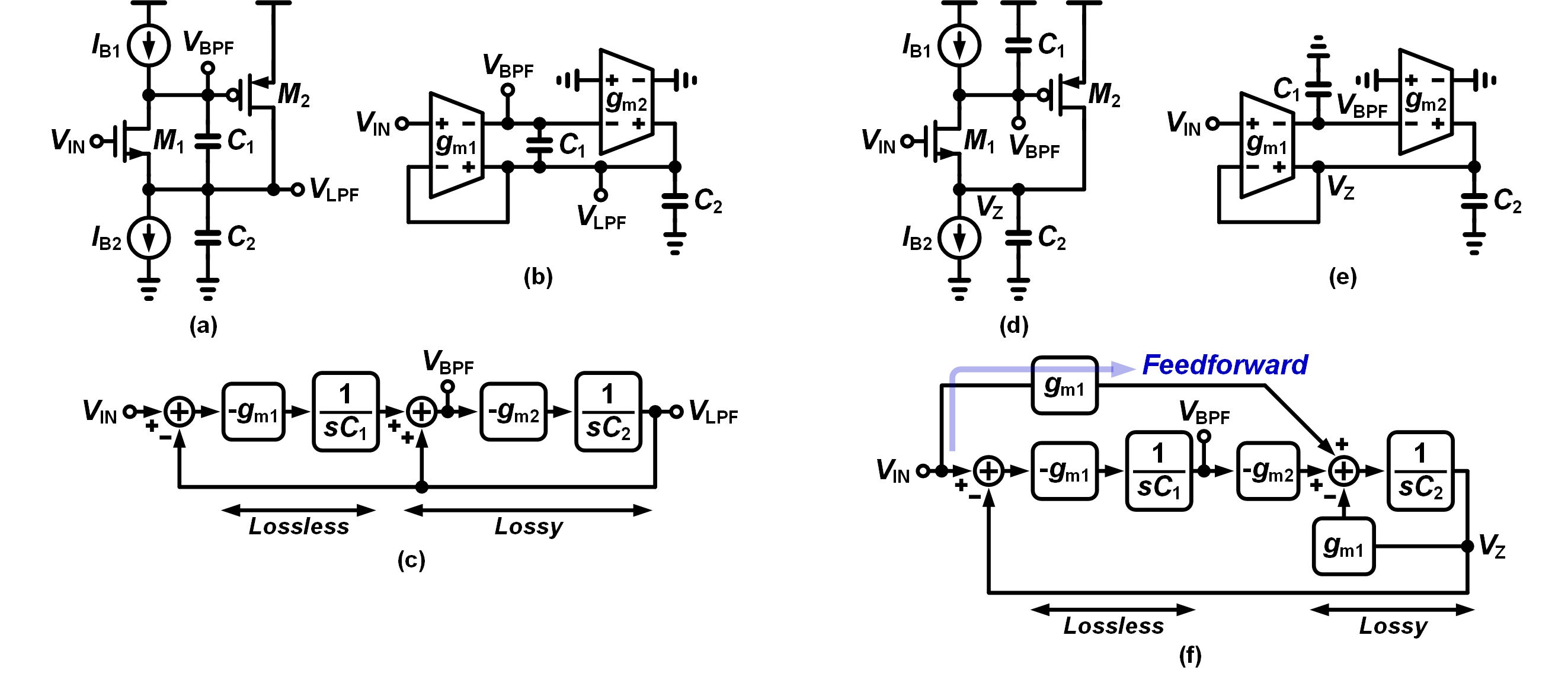}
    \caption{(a) Transistor-level schematic, (b) {$g_\text{m}{C}$} equivalent, and (c) small-signal diagram of type-I SSF \cite{Matteis2015SSF} and (d) transistor-level schematic, (e) {$g_\text{m}{C}$} equivalent, and (f) small-signal diagram of type-II SSF \cite{Yang2019VAD}.}\label{fig:SSF}
    \end{center}
    \vspace{-5mm}
\end{figure*}
\begin{figure*}[!t] \begin{gather}
    H_\text{SSF-LPF-I}(s)
    =\frac{\displaystyle \frac{g_\text{m1}g_\text{m2}}{C_{1}C_{2}}}
    {\displaystyle s^{2}+s\frac{g_\text{m2}}{C_{2}}
    +\frac{g_\text{m1}g_\text{m2}}{C_{1}C_{2}}}\quad
    H_\text{SSF-BPF-I}(s)=\frac{\displaystyle -s\frac{g_\text{m1}}{C_{1}}}
    {\displaystyle s^{2}+s\frac{g_\text{m2}}{C_{2}}
    +\frac{g_\text{m1}g_\text{m2}}{C_{1}C_{2}}}\qquad
    \omega_{0}=\sqrt{\frac{g_\text{m1}g_\text{m2}}{C_{1}C_{2}}}\quad
    Q=\sqrt{\frac{g_\text{m1}C_{2}}{g_\text{m2}C_{1}}}\label{eq:SSF-TypeI}\\
    H_\text{SSF-BPF-II}(s)=\frac{\displaystyle -s\frac{g_\text{m1}}{C_{1}}}
    {\displaystyle s^{2}+s\frac{g_\text{m1}}{C_{2}}
    +\frac{g_\text{m1}g_\text{m2}}{C_{1}C_{2}}}\qquad
    \omega_{0}=\sqrt{\frac{g_\text{m1}g_\text{m2}}{C_{1}C_{2}}}\quad
    Q=\sqrt{\frac{g_\text{m2}C_{2}}{g_\text{m1}C_{1}}}\label{eq:SSF-TypeII}
\end{gather} \end{figure*}

A clear advantage of the \ac{SF}-based filters over the \ac{OTA}-based is its minimal number of parasitic poles. As shown in Fig.\,\,\ref{fig:XSF-LPF}(a), the \ac{XSF} filter exploits every node with the circuit as a source for pole synthesis while the \ac{OTA}-based circuit does not. For instance, the mirror pole in the \ac{OTA} acts as a non-dominant pole thereby necessitating additional power dissipation to uphold the same bandwidth as in the \ac{SF} filter. A parallel filter array adopting the \ac{XSF}-based \ac{BPF} was implemented in \cite{Yang2016Cochlea}; and was used in an environmental sound classification task \cite{ceolini2019audio} and a 2-class speech versus noise task \cite{Kiselev2022AGC} with an \ac{FPGA} environment.

\begin{figure}[t]
    \begin{center}
    \includegraphics[width=\columnwidth]{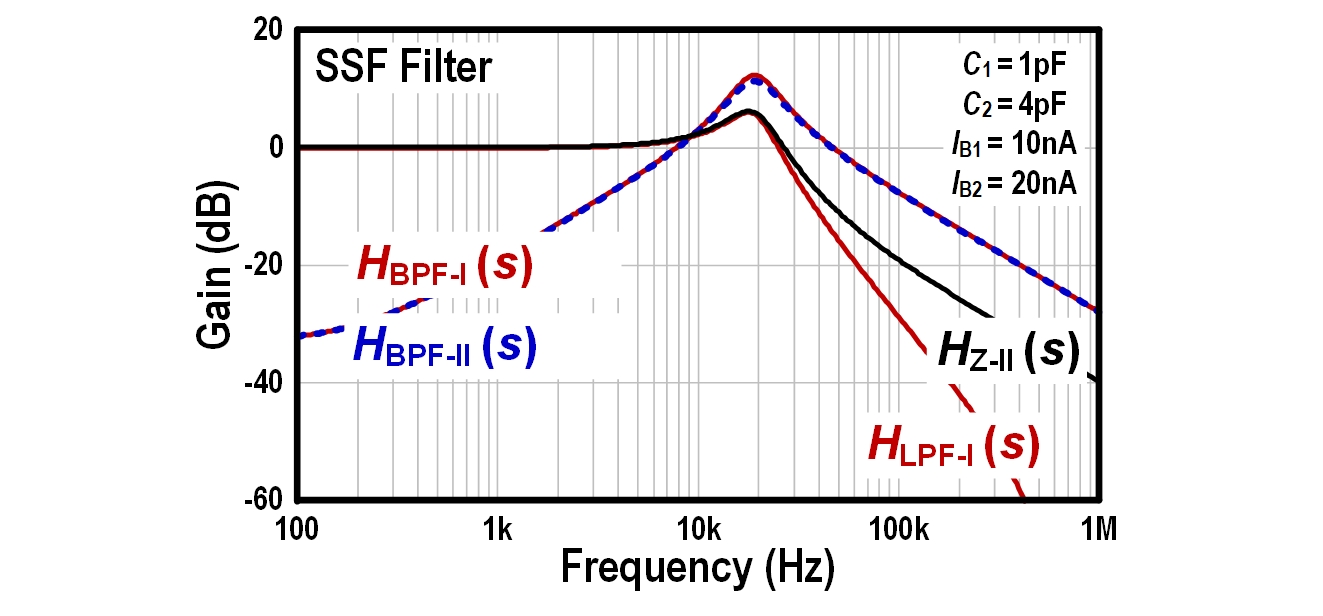}
    \caption{Simulated frequency responses of the \ac{SSF} filter.}\label{fig:SSF-sim}
    \end{center}
    \vspace{-3mm}
\end{figure}

\subsection{Super Source-Follower (Second-Order LPF/BPF)}\label{subsec:SSF}

Fig.\,\,\ref{fig:SSF} shows a schematic, {$g_\text{m}{C}$} equivalent, and block diagram of the super source-follower (SSF)-based filter circuits. As Fig.\,\,\ref{fig:SSF}(a) and (d) show, the \ac{SSF} filter can be categorized into two different types depending on how the $C_{1}$ capacitor is connected: (1) $C_{1}$ bootstraps $V_\text{BPF}$ and $V_\text{LPF}$ in type-I; (2) $C_{1}$ is connected to $V_\text{BPF}$ but its other plate is shunted to GND in type-II. In both types, the basic structure follows the lossless-first two-integrator-loop topology as discussed in Fig.\,\,\ref{fig:biquad_lossless-lossy2}(a). However, the type-II \ac{SSF} incorporates a feedforward $g_\text{m1}$ path. The transfer functions and filter parameters of type-I and type-II \ac{SSF} filters are summarized in (\ref{eq:SSF-TypeI}, \ref{eq:SSF-TypeII}). Interestingly, both \ac{SSF} types can be deployed as a \ac{BPF} without requiring any external subtractor in contrast to the \ac{OTA}-based (see Fig.\,\,\ref{fig:OTA-based-BPF}) and \ac{XSF}-based (see Fig.\,\,\ref{fig:XSF-BPF}) \acp{BPF}. For the type-I \ac{SSF}, this is because $C_{1}$ bootstraps $V_\text{LPF}$ and $V_\text{BPF}$ within the lossless integral operation. In other words, the resulting small-signal voltage difference, i.e., $-g_\text{m1}(v_\text{IN}-v_\text{LPF})sC_{1}$, is generated \textit{on top of} the $V_\text{LPF}$ node. In effect, the $V_\text{BPF}$ is located after the second subtractor which receives $V_\text{LPF}$ as an operand, thereby $V_\text{BPF}$ corresponds to $V_\text{Y}$ in Fig.\,\,\ref{fig:biquad_lossless-lossy2}(a). Note that the small-signal current generated from the $g_\text{m2}$ transconductor does not contribute to the voltage difference over the $C_{1}$ capacitor since the source and sink ports of the $g_\text{m1}$ transconductor are all tied to both plates of the $C_{1}$ capacitor. Assuming sufficiently large output impedance $r_\text{o}$ of the transistors, the small-signal current generated from the $M_{1}$ transistor flows entirely into the $C_{1}$ capacitor, resulting that the generated small-signal current is \textit{trapped} within the $M_{1}\text{-}C_{1}$ loop (used in (\ref{eq:SSF-TypeI-Nodal1})). A similar phenomenon can be found in the noise contribution of cascode devices (see Chapter\,7.4.4 in \cite{razavi2001design}). A nodal analysis for calculating (\ref{eq:SSF-TypeI}) according to Fig.\,\,\ref{fig:SSF}(b) is given as below.
\begin{gather}
    g_\text{m1}(v_\text{IN}-v_\text{LPF})\frac{1}{sC_{1}}+v_\text{BPF}=v_\text{LPF}\\
    v_\text{LPF}=-g_\text{m2}v_\text{BPF}\frac{1}{sC_\text{2}}\label{eq:SSF-TypeI-Nodal1}
\end{gather}

For type-II \ac{SSF}, the key enabler for achieving the \ac{BPF} response is the $g_\text{m1}$ feedforward path. As discussed in (\ref{eq:biquad_BPF_x_w1w2}) and Fig.\,\,\ref{fig:biquad_lossless-lossy2}(b), $V_\text{X}$ within the lossless-first two-integrator-loop topology has a lossy low-frequency behavior. However, the gain path from $V_\text{IN}$ to $V_\text{BPF}$ for calculating the transfer function $H_\text{BPF}(s)$ includes two different paths: (1) direct path $-g_\text{m1}/sC_{1}$; (2) loop around path $g_\text{m1}H_\text{Lossy}(s)g_\text{m1}/sC_{1}$ where $H_\text{Lossy}(s)=1/(g_\text{m1}+sC_{2})$. Since both paths have different polarities, they cancel out each other so that the lossy term in the numerator of (\ref{eq:biquad_BPF_x_w1w2}), i.e., $\omega_{1}\omega_{2}$, is eliminated. A nodal analysis for calculating (\ref{eq:SSF-TypeII}) according to Fig.\,\,\ref{fig:SSF}(e) is given as below.
\begin{gather}
    v_\text{BPF}=-g_\text{m1}(v_\text{IN}-v_\text{Z})\frac{1}{sC_\text{1}}\\
    v_\text{Z}=[g_\text{m1}(v_\text{IN}-v_\text{Z})+(-g_\text{m2}v_\text{BPF})]\frac{1}{sC_\text{2}}
\end{gather}

Note that unlike type-I \ac{SSF}, the remaining node ($V_\text{Z}$) does not show a second-order \ac{LPF} response. The transfer function $H_\text{Z-II}(s)$ of the type-II \ac{SSF} is given as below.
\begin{gather}\label{eq:SSF-TypeII-Z}
    H_\text{Z-II}(s)=\frac{\displaystyle
    \frac{g_\text{m1}}{C_{1}C_{2}}(g_\text{m2}+sC_{1})}
    {\displaystyle s^{2}+s\frac{g_\text{m1}}{C_{2}}
    +\frac{g_\text{m1}g_\text{m2}}{C_{1}C_{2}}}
\end{gather}

Since the transfer function has 1 pole in the numerator and 2 poles in the denominator, similarly to $H_\text{X}(s)$ in (\ref{eq:biquad_BPF_x_w1w2}), it effectively shows a first-order low-pass response.

\begin{figure*}[t]
    \begin{center}
    \includegraphics[width=\textwidth]{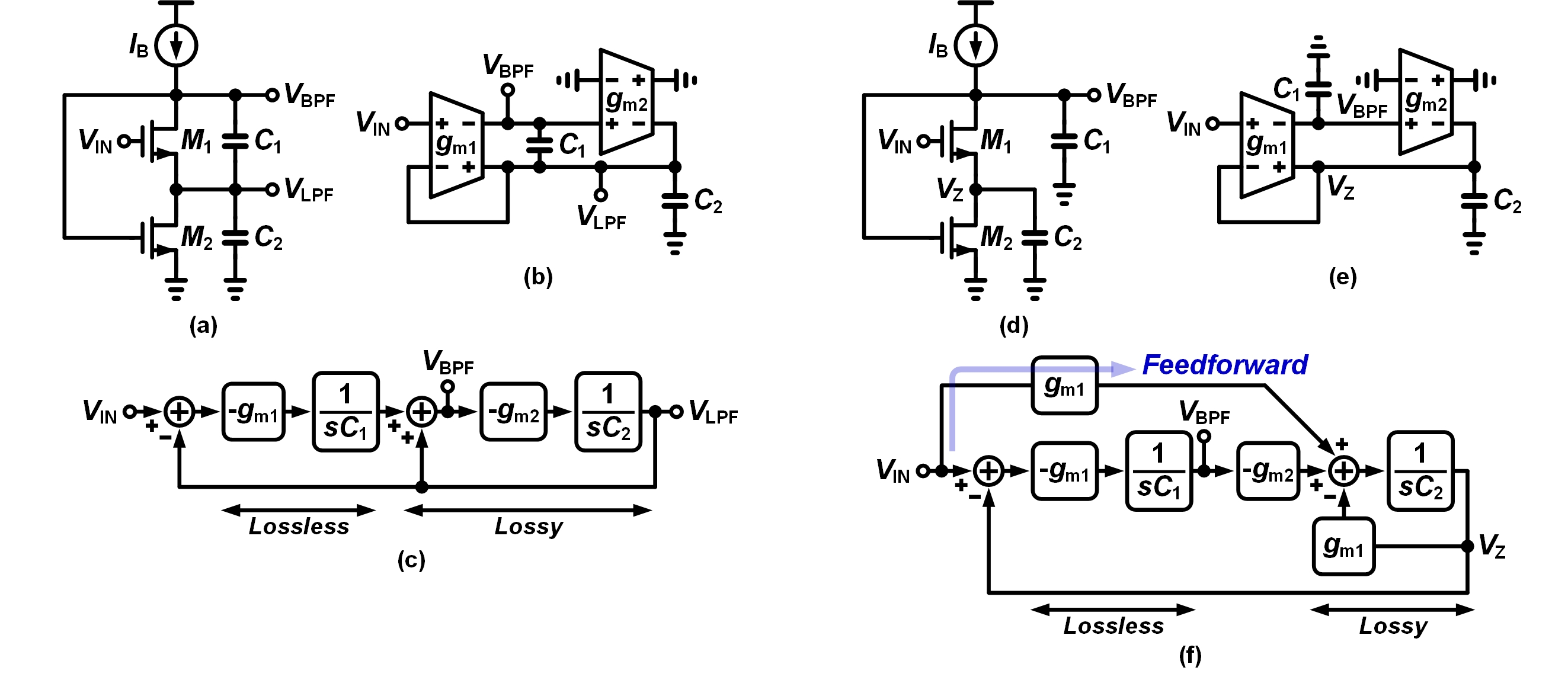}
    \caption{(a) Transistor-level schematic, (b) {$g_\text{m}{C}$} equivalent, and (c) small-signal diagram of the type-I flipped voltage follower \cite{Matteis2017FVF} and (d) transistor-level schematic, (e) {$g_\text{m}{C}$} equivalent, and (f) small-signal diagram of the type-II flipped voltage follower \cite{Yang2021VAD}.}\label{fig:FVF}
    \end{center}
    \vspace{-5mm}
\end{figure*}
\begin{figure*}[!t] \begin{gather}
    H_\text{FVF-LPF-I}(s)
    =\frac{\displaystyle \frac{g_\text{m1}g_\text{m2}}{C_{1}C_{2}}}
    {\displaystyle s^{2}+s\frac{g_\text{m2}}{C_{2}}
    +\frac{g_\text{m1}g_\text{m2}}{C_{1}C_{2}}}\quad
    H_\text{FVF-BPF-I}(s)=\frac{\displaystyle -s\frac{g_\text{m1}}{C_{1}}}
    {\displaystyle s^{2}+s\frac{g_\text{m2}}{C_{2}}
    +\frac{g_\text{m1}g_\text{m2}}{C_{1}C_{2}}}\qquad
    \omega_{0}=\sqrt{\frac{g_\text{m1}g_\text{m2}}{C_{1}C_{2}}}\quad
    Q=\sqrt{\frac{g_\text{m1}C_{2}}{g_\text{m2}C_{1}}}\label{eq:FVF-TypeI}\\
    H_\text{FVF-BPF-II}(s)=\frac{\displaystyle -s\frac{g_\text{m1}}{C_{1}}}
    {\displaystyle s^{2}+s\frac{g_\text{m1}}{C_{2}}
    +\frac{g_\text{m1}g_\text{m2}}{C_{1}C_{2}}}\qquad
    \omega_{0}=\sqrt{\frac{g_\text{m1}g_\text{m2}}{C_{1}C_{2}}}\quad
    Q=\sqrt{\frac{g_\text{m2}C_{2}}{g_\text{m1}C_{1}}}\label{eq:FVF-TypeII}
\end{gather} \end{figure*}

\begin{figure}[t]
    \begin{center}
    \includegraphics[width=\columnwidth]{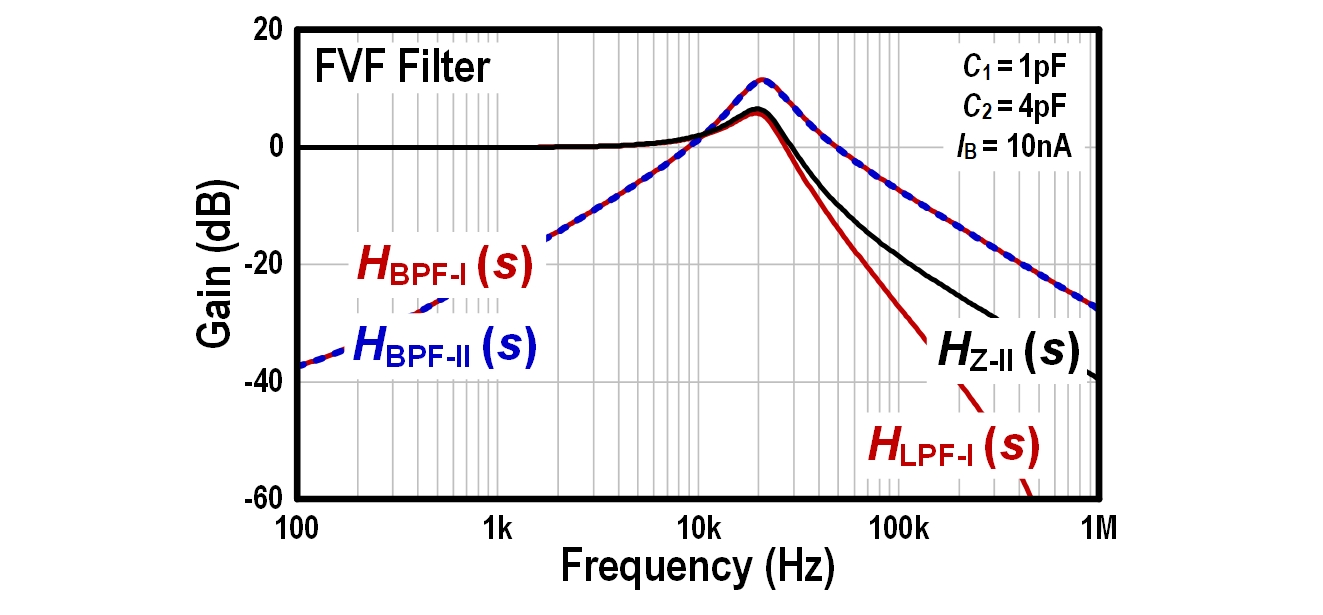}
    \caption{Simulated frequency responses of \ac{FVF} filter.}\label{fig:FVF-sim}
    \end{center}
    \vspace{-5mm}
\end{figure}

Our analysis on the type-I and type-II \ac{SSF} filters are verified with an AC simulation as shown in Fig.\,\,\ref{fig:SSF-sim}. We set $I_\text{B2}=2\times I_\text{B1}=20\,\text{nA}$ such that the same bias currents are distributed to $M_{1}$/$M_{2}$ transistors. Therefore, the transconductances for both transistors are closely set. $g_\text{m1}=252.8\,\text{nS}$ and $g_\text{m2}=227.3\,\text{nS}$. Note that we keep the capacitance ratio as same as the \ac{XSF} filter simulation shown in Fig.\,\,\ref{fig:XSF-LPF-sim}, thereby $Q$ can be designed as $2$. The other simulation setups are the same as mentioned in Section~\ref{subsec:CCSF-LPF}. As predicted in (\ref{eq:SSF-TypeI}, \ref{eq:SSF-TypeII}), Fig.\,\,\ref{fig:SSF-sim} shows that both type-I and type-II \ac{SSF} filters when they are probed at $V_\text{BPF}$ have \ac{BPF} responses. Their peak gains are $12.14\,\text{dB}$ and $11.28\,\text{dB}$ for type-I and type-II cases respectively, and close to our estimated value of $Q^{2}=4=12.04\,\text{dB}$, which is also described in the \ac{XSF} filter simulation. The simulated peak gain values of the two filters are different because the theoretical $Q$ equations are different as given by (\ref{eq:SSF-TypeI}, \ref{eq:SSF-TypeII}) considering slightly different $g_\text{m}$ values for $M_{1}$/$M_{2}$. A second-order low-pass roll-off is observed with $H_\text{LPF-I}(s)$ and a first-order roll-off with $H_\text{Z-II}(s)$.
The center frequencies of type-I/type-II \ac{SSF} \acp{BPF} can be estimated using (\ref{eq:SSF-TypeI}, \ref{eq:SSF-TypeII}) as below.
\begin{equation}
f_{0}=\frac{\omega_{0}}{2\pi}
=\frac{\sqrt{252.8\,\text{nS}\times 227.3\,\text{nS}}}
{2\pi\sqrt{1\,\text{pF}\times 4\,\text{pF}}}
=19.08\,\text{kHz}
\end{equation}
which makes a close agreement with $19.05\,\text{kHz}$ from the simulation result in {Fig.\,\,\ref{fig:SSF-sim}.} The type-II \ac{SSF}-based \ac{BPF} was implemented within a channel of a parallel filter bank feature extractor in \cite{Yang2019VAD}. Together with an on-chip \ac{MLP} classifier, it implemented a \ac{VAD}.

\subsection{Flipped Voltage Follower (Second-Order LPF/BPF)}\label{subsec:FVF}

Fig.\,\,\ref{fig:FVF} shows a schematic, {$g_\text{m}{C}$} equivalent, and block diagram of the flipped voltage follower (FVF)-based filter circuits \cite{Carvajal2005FVF}. The \ac{FVF} circuit has been actively used as a core building block in various analog circuits to name, such as \ac{LPF} \cite{Matteis2017FVF}, low-dropout (LDO) regulator \cite{Man2008FVF-LDO, Guo2010LDO, Gweon2019LDO}, bio-signal amplifier \cite{Kim2017BioZ, Kim2020BioZ, Ha2019BioZ, Xu2018NIRS-EEG-EIT, Lee2020EIT, Helleputte2012IA}, and current driver \cite{Kim2022BioZ, Kim2020BioZVLSI}. As of the case in the \ac{SSF} filters, the \ac{FVF} filters are also categorized into two types according to the connection methods of $C_{1}$ capacitor. Interestingly, the {$g_\text{m}{C}$} equivalent circuits of the \ac{SSF} and the \ac{FVF} are the same except for the input polarity of $g_\text{m2}$ transconductor (pFET gate in \ac{SSF}, nFET gate in \ac{FVF}). Since the inversion characteristic of $g_\text{m2}$ transconductor still does not change, resultant block diagrams of the \ac{SSF} and \ac{FVF} are exactly same regardless of whether they are type-I or type-II. Therefore, the transfer functions of \ac{FVF} filters described in (\ref{eq:FVF-TypeI}), (\ref{eq:FVF-TypeII}) are also the same as of the \ac{SSF} filters (\ref{eq:SSF-TypeI}), (\ref{eq:SSF-TypeII}).

Fig.\,\,\ref{fig:FVF-sim} shows an AC simulation result of the type-I and type-II \ac{FVF} filter circuits. We set $I_\text{B}=10\,\text{nA}$ and the resulting transconductances are $g_\text{m1}=\text{262.4}\,\text{nS}$ and $g_\text{m2}=\text{262.5}\,\text{nS}$. The other simulation setups are the same as \ac{XSF} in Section~\ref{subsec:CCSF-LPF} and \ac{SSF} in Section~\ref{subsec:SSF}. As expected, the transfer curves for $H_\text{LPF-I}$, $H_\text{BPF-I}$, $H_\text{BPF-II}$, and $H_\text{Z-II}$ show the same characteristic as we observed in the \ac{SSF} simulation result (Fig.\,\,\ref{fig:SSF-sim}). The peak gains of type-I and type-II \acp{BPF} are $11.39\,\text{dB}$ and $11.35\,\text{dB}$ respectively, where their center frequency is observed as $20.89\,\text{kHz}$. The simulated filter parameters are in close agreement with our estimation using (\ref{eq:FVF-TypeI}) and (\ref{eq:FVF-TypeII}), $Q^{2}=4=12.04\,\text{dB}$ and
\begin{equation}
f_{0}=\frac{\omega_{0}}{2\pi}
=\frac{\sqrt{262.4\,\text{nS}\times 262.5\,\text{nS}}}
{2\pi\sqrt{1\,\text{pF}\times 4\,\text{pF}}}
=20.86\,\text{kHz}
\end{equation}

A significant advantage of the \ac{FVF} over the \ac{SSF} is its power efficiency in terms of bandwidth. More specifically, as shown in Fig.\,\,\ref{fig:SSF-sim} and Fig.\,\,\ref{fig:FVF-sim}, the \ac{FVF} consumes 2$\times$ less current ($I_\text{B}=10\,\text{nA}$) while the \ac{SSF} consumes $I_\text{B}=20\,\text{nA}$ to achieve $\omega_{0}=20\,\text{kHz}$ for $C_{2}=4\times C_{1}=4\,\text{pF}$. This is because the tail current $I_\text{B2}$ in the \ac{SSF} is divided into two different bias currents, $I_\text{B1}$ and $I_\text{B2}-I_\text{B1}$, for $g_\text{m1}$ and $g_\text{m2}$ respectively. On the contrary, the \ac{FVF} exploits its inherent current reusing nature to save the power. More importantly, the \ac{FVF} circuit offers better matching over the \ac{SSF}, not merely because of the single branch biasing, but also of the same transistor type. For example, $M_{1}$ and $M_{2}$ transistors are all nFET in the \ac{FVF} while it is not the case in the \ac{SSF}. As a result, the \ac{FVF} circuit is more robust over process variation because it is difficult to match different transistor types especially in regards to the shallow trench isolation (STI) and well proximity effect (WPE) \cite{Drennan2006WPE}. The type-II \ac{FVF}-based \ac{BPF} was implemented as a parallel audio \ac{FEx} for a VAD \ac{IC} in \cite{Yang2021VAD}.


\begin{table*}[ht]
\centering
\caption{Summary of audio FEx and their usage for edge audio intelligence}\label{table:summary}
\begin{tabular}{|cccccccc||ccc|}
\hline
\multicolumn{8}{|c||}{Audio Feature Extractor (FEx)}
& \multicolumn{3}{c|}{Audio Inference Task}
\\ \hline
\multicolumn{1}{|c|}{Paper} &           
\multicolumn{3}{c|}{Filter Type} &      
\multicolumn{1}{c|}{\begin{tabular}[c]{@{}c@{}}
Process\\(nm)\end{tabular}} &           
\multicolumn{1}{c|}{Filter Bank} &      
\multicolumn{1}{c|}{Power} &            
\begin{tabular}[c]{@{}c@{}}
Area\\(mm\textsuperscript{2})\end{tabular}
&                                       
\multicolumn{1}{c|}{\begin{tabular}[c]{@{}c@{}}
Task\\(\# of Classes)\end{tabular}} &   
\multicolumn{1}{c|}{Dataset} &          
\multicolumn{1}{c|}{Classifier}         
\\ \hline
\multicolumn{1}{|c|}{\begin{tabular}[c]{@{}c@{}}
TBioCAS\\2014 \cite{liu2014cochlea}
\end{tabular}} &                        
\multicolumn{1}{c|}{\multirow{17}{*}{\begin{tabular}[c]{@{}c@{}}
{CT}\end{tabular}}} &       
\multicolumn{1}{c|}{\multirow{14}{*}{\begin{tabular}[c]{@{}c@{}}
Analog\\Voltage\\({$g_\text{m}{C}$})
\end{tabular}}} &                       
\multicolumn{1}{c|}{OTA-1} & 
\multicolumn{1}{c|}{350} &              
\multicolumn{1}{c|}{\begin{tabular}[c]{@{}c@{}}
2$\times$64$\times$4 Ch\\50\,Hz-50\,kHz
\end{tabular}} &                        
\multicolumn{1}{c|}{14\,mW
{\color{Maroon}\textsuperscript{A}}} &  
13.74 &                                 
\multicolumn{1}{c|}{\begin{tabular}[c]{@{}c@{}}
{ASR\textsuperscript{\color{Maroon}D}}\\
(12)\end{tabular}} &                     
\multicolumn{1}{c|}{TIDIGITS} &          
\multicolumn{1}{c|}{\begin{tabular}[c]{@{}c@{}}
FPGA\\RNN
\cite{Gao2019TIDIGITS}
\end{tabular}}                          
\\ \cline{1-1} \cline{4-11}
\multicolumn{1}{|c|}{\begin{tabular}[c]{@{}c@{}}
JSSC\\2016 \cite{Yang2016Cochlea}
\end{tabular}} &                        
\multicolumn{1}{c|}{} &                 
\multicolumn{1}{c|}{} &                 
\multicolumn{1}{c|}{XSF-1} & 
\multicolumn{1}{c|}{180} &              
\multicolumn{1}{c|}{\begin{tabular}[c]{@{}c@{}}
2$\times$64 Ch\\8\,Hz-20\,kHz
\end{tabular}} &                        
\multicolumn{1}{c|}{55\,$\mu$W} &       
33.28 &                                 
\multicolumn{1}{c|}{\begin{tabular}[c]{@{}c@{}}
Speech\\vs Noise (2)\end{tabular}} &    
\multicolumn{1}{c|}{\begin{tabular}[c]{@{}c@{}}
TIMIT\\MS-SNSD\\MUSAN\end{tabular}} &   
\multicolumn{1}{c|}{\begin{tabular}[c]{@{}c@{}}
Software\\MLP
\cite{Kiselev2022AGC}
\end{tabular}}                          
\\ \cline{1-1} \cline{4-11}
\multicolumn{1}{|c|}{\begin{tabular}[c]{@{}c@{}}
JSSC\\2016 \cite{Badami2016VAD}
\end{tabular}} &                        
\multicolumn{1}{c|}{} &                 
\multicolumn{1}{c|}{} &                 
\multicolumn{1}{c|}{OTA-2} & 
\multicolumn{1}{c|}{90} &               
\multicolumn{1}{c|}{\begin{tabular}[c]{@{}c@{}}
16 Ch\\75\,Hz-5\,kHz
\end{tabular}} &                        
\multicolumn{1}{c|}{6\,$\mu$W} &        
2 &                                     
\multicolumn{1}{c|}{VAD (2)} &          
\multicolumn{1}{c|}{NOIZEUS} &          
\multicolumn{1}{c|}{\begin{tabular}[c]{@{}c@{}}
On-Chip\\Decision Tree\end{tabular}}    
\\ \cline{1-1} \cline{4-11}
\multicolumn{1}{|c|}{\begin{tabular}[c]{@{}c@{}}
ISCAS\\2017 \cite{Shah2017FPAA}
\end{tabular}} &                        
\multicolumn{1}{c|}{} &                 
\multicolumn{1}{c|}{} &                 
\multicolumn{1}{c|}{OTA-2} & 
\multicolumn{1}{c|}{350} &  
\multicolumn{1}{c|}{\begin{tabular}[c]{@{}c@{}}
12 Ch\\100\,Hz-5\,kHz
\end{tabular}} &                        
\multicolumn{1}{c|}{153.8\,$\mu$W} &        
{-} &                       
\multicolumn{1}{c|}{VAD (2)} &          
\multicolumn{1}{c|}{TIMIT} &          
\multicolumn{1}{c|}{\begin{tabular}[c]{@{}c@{}}
On-Chip\\WTA\end{tabular}}    
\\ \cline{1-1} \cline{4-11}
\multicolumn{1}{|c|}{\begin{tabular}[c]{@{}c@{}}
JSSC\\2019 \cite{Yang2019VAD}
\end{tabular}} &                        
\multicolumn{1}{c|}{} &                 
\multicolumn{1}{c|}{} &                 
\multicolumn{1}{c|}{SSF-1} & 
\multicolumn{1}{c|}{180} &              
\multicolumn{1}{c|}{\begin{tabular}[c]{@{}c@{}}
16 Ch\\100\,Hz-5\,kHz
\end{tabular}} &                        
\multicolumn{1}{c|}{380\,nW} &          
1.6 &                                   
\multicolumn{1}{c|}{VAD (2)} &          
\multicolumn{1}{c|}{\begin{tabular}[c]{@{}c@{}}
Aurora4\\DEMAND\end{tabular}} &         
\multicolumn{1}{c|}{On-Chip MLP}        
\\ \cline{1-1} \cline{4-11}
\multicolumn{1}{|c|}{\begin{tabular}[c]{@{}c@{}}
JSSC\\2021 \cite{Yang2021VAD}
\end{tabular}} &                        
\multicolumn{1}{c|}{} &                 
\multicolumn{1}{c|}{} &                 
\multicolumn{1}{c|}{FVF-1} & 
\multicolumn{1}{c|}{65} &               
\multicolumn{1}{c|}{\begin{tabular}[c]{@{}c@{}}
16 Ch\\100\,Hz-5\,kHz
\end{tabular}} &                        
\multicolumn{1}{c|}{52\,nW} &           
0.9 &                                   
\multicolumn{1}{c|}{\begin{tabular}[c]{@{}c@{}}
VAD (2)\\KWS (2)\end{tabular}} &        
\multicolumn{1}{c|}{\begin{tabular}[c]{@{}c@{}}
Aurora4\\DEMAND\\GSCD\end{tabular}} &   
\multicolumn{1}{c|}{\begin{tabular}[c]{@{}c@{}}
On-Chip MLP\\(VAD)\\Software CNN\\(KWS)
\end{tabular}}                          
\\ \cline{1-1} \cline{4-11}
\multicolumn{1}{|c|}{\begin{tabular}[c]{@{}c@{}}
ISSCC\\2021 \cite{Wang2021KWS}
\end{tabular}} &                        
\multicolumn{1}{c|}{} &                 
\multicolumn{1}{c|}{} &                 
\multicolumn{1}{c|}{-} &    
\multicolumn{1}{c|}{65} &               
\multicolumn{1}{c|}{\begin{tabular}[c]{@{}c@{}}
16 Ch\\-
\end{tabular}} &                        
\multicolumn{1}{c|}{109\,nW} &          
0.72 &                                  
\multicolumn{1}{c|}{KWS (5)} &          
\multicolumn{1}{c|}{\begin{tabular}[c]{@{}c@{}}
HeySnips\\GSCD\end{tabular}} &          
\multicolumn{1}{c|}{On-Chip MLP}        
\\ \cline{1-1} \cline{3-11}
\multicolumn{1}{|c|}{\begin{tabular}[c]{@{}c@{}}
ISSCC\\2022 \cite{Kim2022KWS}
\end{tabular}} &                        
\multicolumn{1}{c|}{} &                 
\multicolumn{1}{c|}{\begin{tabular}[c]{@{}c@{}}
Analog\\Time\\(OSC)\end{tabular}} &     
\multicolumn{1}{c|}{OSC-2} & 
\multicolumn{1}{c|}{65} &               
\multicolumn{1}{c|}{\begin{tabular}[c]{@{}c@{}}
16 Ch\\111\,Hz-10.4\,kHz
\end{tabular}} &                        
\multicolumn{1}{c|}{9.3\,$\mu$W} &      
1.6 &                                   
\multicolumn{1}{c|}{KWS (12)} &         
\multicolumn{1}{c|}{GSCD} &             
\multicolumn{1}{c|}{On-Chip RNN}        
\\ \hline
\multicolumn{1}{|c|}{\begin{tabular}[c]{@{}c@{}}
JSSC\\2019 \cite{Oh2019VAD}
\end{tabular}} &                        
\multicolumn{1}{c|}{\multirow{5}{*}{\begin{tabular}[c]{@{}c@{}}
{DT}\end{tabular}}} &       
\multicolumn{1}{c|}{\begin{tabular}[c]{@{}c@{}}
Analog\\Voltage\\(Mixer)
\end{tabular}} &                        
\multicolumn{1}{c|}{-} &    
\multicolumn{1}{c|}{180} &              
\multicolumn{1}{c|}{\begin{tabular}[c]{@{}c@{}}
16-48 Ch\\75\,Hz-4\,kHz
\end{tabular}} &                        
\multicolumn{1}{c|}{60\,nW} &           
0.55{\color{Maroon}
\textsuperscript{B}} &                  
\multicolumn{1}{c|}{VAD (2)} &          
\multicolumn{1}{c|}{\begin{tabular}[c]{@{}c@{}}
LibriSpeech\\NOISEX-92\end{tabular}} &  
\multicolumn{1}{c|}{On-Chip MLP}        
\\ \cline{1-1} \cline{3-11}
\multicolumn{1}{|c|}{\begin{tabular}[c]{@{}c@{}}
TCAS-I\\2021 \cite{Villamizar2021SC}
\end{tabular}} &                        
\multicolumn{1}{c|}{} &                 
\multicolumn{1}{c|}{\multirow{3}{*}{\begin{tabular}[c]{@{}c@{}}
Analog\\Voltage\\(SC)\end{tabular}}} &  
\multicolumn{1}{c|}{-} &    
\multicolumn{1}{c|}{130} &              
\multicolumn{1}{c|}{\begin{tabular}[c]{@{}c@{}}
32 Ch\\30\,Hz-8\,kHz
\end{tabular}} &                        
\multicolumn{1}{c|}{800\,nW} &          
0.79 &                                  
\multicolumn{1}{c|}{KWS (12)} &         
\multicolumn{1}{c|}{GSCD} &             
\multicolumn{1}{c|}{Software RNN}       
\\ \cline{1-1} \cline{4-11}
\multicolumn{1}{|c|}{\begin{tabular}[c]{@{}c@{}}
SSC-L\\2022 \cite{Fuketa2022SC}
\end{tabular}} &                        
\multicolumn{1}{c|}{} &                 
\multicolumn{1}{c|}{} &                 
\multicolumn{1}{c|}{-} &    
\multicolumn{1}{c|}{65} &               
\multicolumn{1}{c|}{\begin{tabular}[c]{@{}c@{}}
6 Ch\\20\,Hz-4\,kHz
\end{tabular}} &                        
\multicolumn{1}{c|}{150\,nW} &          
0.84 &                                  
\multicolumn{1}{c|}{KWS (3)} &          
\multicolumn{1}{c|}{GSCD} &             
\multicolumn{1}{c|}{Software MLP}       
\\ \hline \hline
\multicolumn{1}{|c|}{\begin{tabular}[c]{@{}c@{}}
TCAS-I\\2019 \cite{Zheng2019Speech}
\end{tabular}} &                        
\multicolumn{2}{c|}{\multirow{5}{*}{\begin{tabular}[c]{@{}c@{}}
Digital\\(FFT)\end{tabular}}} &         
\multicolumn{1}{c|}{-} &    
\multicolumn{1}{c|}{28} &               
\multicolumn{1}{c|}{\begin{tabular}[c]{@{}c@{}}
40 Ch\\0\,Hz-8\,kHz
\end{tabular}} &                        
\multicolumn{1}{c|}{9.19\,$\mu$W{\color{Maroon}
\textsuperscript{C}}} &                 
0.08{\color{Maroon}
\textsuperscript{C}} &                  
\multicolumn{1}{c|}{KWS (2)} &          
\multicolumn{1}{c|}{TIDIGITS} &         
\multicolumn{1}{c|}{On-Chip CNN}        
\\ \cline{1-1} \cline{4-11}
\multicolumn{1}{|c|}{\begin{tabular}[c]{@{}c@{}}
VLSI\\2019 \cite{giraldo2019kws}
\end{tabular}} &                        
\multicolumn{2}{c|}{} &                 
\multicolumn{1}{c|}{-} &    
\multicolumn{1}{c|}{65} &               
\multicolumn{1}{c|}{\begin{tabular}[c]{@{}c@{}}
20 Ch\\0\,Hz-8\,kHz
\end{tabular}} &                        
\multicolumn{1}{c|}{7.33\,$\mu$W{\color{Maroon}
\textsuperscript{C}}} &                 
0.96{\color{Maroon}
\textsuperscript{B}} &                  
\multicolumn{1}{c|}{KWS (12)} &         
\multicolumn{1}{c|}{GSCD} &             
\multicolumn{1}{c|}{On-Chip RNN}        
\\ \cline{1-1} \cline{4-11}
\multicolumn{1}{|c|}{\begin{tabular}[c]{@{}c@{}}
JSSC\\2021 \cite{shan2021kws}
\end{tabular}} &                        
\multicolumn{2}{c|}{} &                 
\multicolumn{1}{c|}{-} &    
\multicolumn{1}{c|}{28} &               
\multicolumn{1}{c|}{\begin{tabular}[c]{@{}c@{}}
10 Ch\\0\,Hz-4\,kHz
\end{tabular}} &                        
\multicolumn{1}{c|}{340\,nW} &          
0.05{\color{Maroon}
\textsuperscript{B}} &                  
\multicolumn{1}{c|}{KWS (2-5)} &        
\multicolumn{1}{c|}{GSCD} &             
\multicolumn{1}{c|}{On-Chip CNN}        
\\ \hline
\multicolumn{11}{|l|}{
OTA-1: \ac{OTA}-based \& lossless-first$\qquad\qquad$
OTA-2: \ac{OTA}-based \& lossy-first$\qquad\qquad$\,\,\,\,\,
XSF-1: \ac{XSF}-based \& lossless-first}
\\
\multicolumn{11}{|l|}{
SSF-1: \ac{SSF}-based \& lossless-first$\qquad\qquad$\,\,\,
FVF-1: \ac{FVF}-based \& lossless-first$\qquad\qquad$
OSC-2: OSC-based \& lossy-first}\\
\multicolumn{11}{|l|}{
{\color{Maroon}\textsuperscript{A}}This number includes power consumption of the microphone pre-amplifier and test circuits.}\\
\multicolumn{11}{|l|}{
{\color{Maroon}\textsuperscript{B}}Estimated from chip photograph.$\qquad$
{\color{Maroon}\textsuperscript{C}}Estimated from power/area breakdown.$\qquad$
{{
\textsuperscript{\color{Maroon}D}}Automatic Speech Recognition.}}
\\ \hline
\end{tabular}
\end{table*}

\section{Summary and Outlook}\label{sec:Summary}

Table\,\,\ref{table:summary} summarizes the audio feature extractor (FEx) \acp{IC} reported for edge \ac{AI} tasks such as \ac{VAD} and \ac{KWS}. {In this table, we only compare those \ac{FEx} \acp{IC} that were validated with fabricated chip measurements but there are other reported \ac{FEx} designs that could also be useful for the edge AI tasks.} To date, both analog and digital \acp{FEx} have been used in audio edge devices. An analog \ac{FEx} can be categorized as a \ac{CT} or \ac{DT} filter while a \ac{CT} filter can be designed using voltage-domain or time-domain circuits. Here, the \textit{Time-Domain} means the signals are processed using \ac{PWM} or \ac{PFM}-based circuits. Note that it is still a \ac{CT} signal which is not sampled by a clock running at a known frequency. For instance, a \ac{TDC}, which is widely used in \acp{PLL} \cite{Hong2012TDC} and \ac{ToF} \cite{Kondo2020TDC} sensors, converts an \textit{Analog} time-domain \ac{PWM} input signal into a sampled and quantized \textit{Digital} output.

The \ac{CT} voltage-domain filter discussed in Section\,\,\ref{subsec:OTA_BPF} is combined with a rectifier and a spike generation stage to form a cochlea channel leading to the multi-channel \ac{DAS} silicon cochlea \cite{liu2014cochlea}. This design has been used in applications such as sound source localization \cite{anumula2018localization,XuMachineHr2018} using the spike timing of the binaural spikes from the \ac{DAS}. It was also used for multi-modal recognition \cite{li2019lip,kiselev2016event}, and keyword spotting using \acp{DNN} \cite{Gao2019TIDIGITS, ceolini2019icassp}. The latest \ac{CT} voltage-domain filter circuits show sub-$\mu$W ultra-low-power consumption \cite{Yang2016Cochlea, Yang2019VAD, Yang2021VAD, Wang2021KWS, Badami2016VAD}, by mainly exploiting the outstanding power efficiency of \ac{SF}-based filters operating in subthreshold as discussed in Section\,\,\ref{sec:SF_filter}. Over a range of \ac{CT} voltage-domain analog filters discussed in this paper, we may conclude that the \ac{FVF}-based second-order filter is the best option to be adopted for implementing edge audio devices{. This is because 1) it benefits from its intrinsic negative feedback as the same case of the \ac{SF}, 2) similar to the \ac{SSF}, it builds a \ac{BPF} without additional subtraction stages, e.g., \ac{XSF} in Fig.~\ref{fig:XSF-BPF}, 3) it has 2$\times$ higher power-efficiency than \ac{SSF} as discussed in Section~\ref{subsec:FVF}, 4) its transconductors are made of only nFETs or pFETs, which can lead to better design compactness and thus better matching than the \ac{SSF}.}

However, as shown in Section\,\,\ref{subsec:sf} with Eq.\,\,(\ref{eq:sf-LPF_loopgain}), the core strength of the \ac{SF}-based filter is the intrinsic feedback within the circuit. Unfortunately, the loop gain starts to degrade as technology scales, leading to a higher output non-linearity assuming the transistor size also scales. In addition, the reduced voltage headroom mainly caused by faster $V_\text{DD}$ scaling than $V_\text{TH}$, is expected to further complicate the analog filter design forcing \ac{IC} designers to bring concessions in circuit performances. To this end, an analog \ac{FEx} that uses the time-domain processing technique is recently reported in \cite{Kim2022KWS-JSSC}. In contrast to the voltage-domain designs, the building blocks of the time-domain processing circuits are based on logic gates and thus it can potentially provide better technology scaling. In addition, it is also expected to be developed towards a fully-synthesizable analog \ac{FEx} where the layout is automatically generated from the \ac{RTL} hardware description language. Examples of synthesizable analog circuits include all-digital phase-locked loop (ADPLL) \cite{deng2015adpll} and ring-oscillator-based $\Delta\Sigma$ \ac{ADC} \cite{li2019vcodsm}.

Although not discussed in this paper, the \ac{DT} voltage-domain filter circuits are also promising candidates. This is because the center frequency of the \ac{BPF} is controlled by the frequency of an external clock, rather than $g_\text{m}$ of the transconductors, therefore  $\omega_{0}$ can be precisely controlled over process, voltage, and temperature (PVT) variations. A chopper-based mixer with a sequentially varying clock frequency and a subsequent \ac{LPF} stage was used in \cite{Oh2019VAD} where its operational principle is similar to that of lock-in amplifiers, also commonly used in bio-impedance sensors \cite{Kim2019BioZ, Ha2019BioZ}. This architecture achieved a 60\,nW ultra-low-power consumption, however, because it sequentially demodulates over the desired frequency band, it cannot perform the filtering operation over its entire frequency range at once. Therefore, this design showed a 512\,ms latency until a set of filtered data is collected such that a frequency-selective feature vector to be available. Alternatively, the \ac{SC} \acp{BPF} were developed in \cite{Villamizar2021SC, Fuketa2022SC} with a parallel filter bank approach. As typically considered in $\Delta\Sigma$ \cite{Pavan2008CTDSM} and \ac{SAR} \cite{Shen2019CT-SAR} \ac{ADC} designs, the synchronous \ac{SC} operation comes with $kT/C$ noise-aliasing due to the \ac{DT} sample-and-hold. This attribute necessitates an anti-aliasing filter and also a buffer stage ahead of the \ac{SC} filter both of which incur additional power and area \cite{Shen2019CT-SAR}, although they were not actually implemented in \cite{Villamizar2021SC, Fuketa2022SC}. With this \ac{DT} sample-and-hold environment, capacitance must be increased to reduce the $kT/C$ noise, but this choice comes with a larger capacitor area, a higher switching power of the \ac{SC} operation, and a higher capacitive driving strength required for the front-end buffer stage which results in higher power consumption. {In fact,} the work in \cite{Villamizar2021SC} adopted high-density and low-leakage ferroelectric capacitors to realize a low silicon area but it is typically unavailable in standard CMOS process \cite{Butzen2016SCPC}. Also note that cascading approach of a \ac{LPF} and a \ac{HPF} to build a \ac{SC}-\ac{BPF}, adopted in \cite{Fuketa2022SC}, exhibited a limited $Q$ factor ($\leq 0.5$) as discussed in Fig.\,\,\ref{fig:lossy_cascade} and Eq.\,\,\ref{eq:lossy_cascade}.

There are approaches to optimize the \ac{FFT}-based digital \ac{FEx} design to reduce the power consumption toward sub-$\mu$W. These designs implemented either a \ac{FFT} \cite{shan2021kws, Zheng2019Speech} or a \ac{DFT} \cite{giraldo2019kws} computation unit and this is typically followed by Mel filtering and logarithmic compression circuits. {These digital \acp{FEx} are easier to be implemented into a silicon chip than analog approaches, as they can be automatically place-and-routed from the \ac{RTL} code such as Verilog. Therefore, they offer simpler designs, shorter implementation time, and easier portability between different process nodes. In the \ac{KWS} \ac{IC} reported in \cite{giraldo2019kws}, the full signal chain starting from the \ac{AFE} consisting of a voltage amplifier and a 10-bit \ac{SAR} \ac{ADC}, to the digital back-end consisting of a \ac{FFT}-based digital \ac{FEx} and a \ac{RNN}-based classifier is implemented on-chip. The \ac{FEx} alone consumed 7.33\,$\mu$W or 40\,\% of the total power (16.1\,$\mu$W).}
{By using a serialized \ac{FFT} approach \cite{shan2021kws}, the \ac{FEx} power is reduced to only 340\,nW,} however, it relied on an off-chip 16-bit \ac{ADC} which incurs additional power and area. Note that the state-of-the-art {15.2 effective number of bits (ENOB)} $\Delta\Sigma$ modulator with a 5\,kHz bandwidth (close to 4\,kHz used in \cite{shan2021kws}) already consumes 4.5\,$\mu$W \cite{murmann2021ADC-Survey, Chandrakumar2018DSM} and this power number did not {include the power of} the decimation filter stage which is an essential building block for the $\Delta\Sigma$ modulators {in eliminating} high-pass shaped quantization noise. Therefore, it should be {emphasized} that the actual power number of the 16-bit \ac{ADC} will be higher than 4.5\,$\mu$W in edge audio devices operated in the real world. {Note that we have not covered digital designs of biquad filters in this manuscript, e.g., the \ac{FPGA}-based cochlea-inspired designs in \cite{Nair2022IoT, Xu2018CAR-FAC}.} 


\section{Conclusion}\label{sec:Conclusion}

This paper introduces an overview of continuous-time analog filters which have been used for audio edge intelligence applications. A unified analysis of second-order voltage-domain filters using the two-integrator-loop interpretation is presented. With a review of several filter architectures ranging from the OTA-based to source-follower-based designs, {$g_\text{m}{C}$} equivalents and small-signal diagrams are summarized. The derived transfer functions are also verified with the transistor-level simulations. We provide a summary of the state-of-the-art audio feature extraction circuits that have been used for edge audio tasks, also with discussions of their current challenges and design advantages. 

{There are a couple of interesting directions for future analysis. One is the gain and stability analysis of cascaded and parallel cochlea filter bank architectures using the different filter designs \cite{mead1989a-vlsi}. The second is for future detailed circuit analysis which considers the impact of mismatch and circuit nonlinearities on the transfer function of the different filter variants. For both studies, one would require the specifications of a common fabrication technology and the choice of power supply voltage and transistor sizes for a fair comparison. However, various studies have shown that deep networks can learn to incorporate circuit nonlinearities and quantizatioin noise if these feature nonidealities are present in the training samples of a network, similar to that carried out in \cite{Yang2019VAD, Kiselev2022AGC}.} 
\section*{Acknowledgment}
The authors would like to thank Tobi Delbruck, Chang Gao, and Sheng Zhou of the Sensors Group at the Institute of Neuroinformatics, for discussions and feedback on the circuit analysis.

\bibliographystyle{IEEEtran}
\bibliography{main}

\begin{IEEEbiography}[{\includegraphics[width=1in,height=1.25in,clip,keepaspectratio]{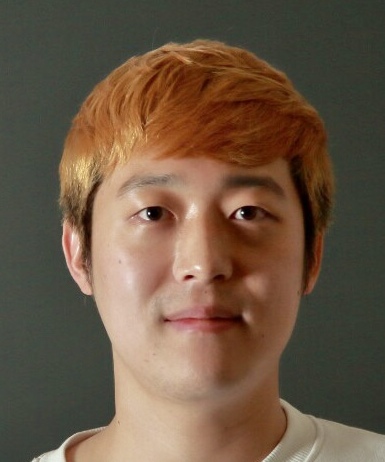}}]{Kwantae Kim}
(Member, IEEE) received the B.S., M.S., and Ph.D. degrees in the School of Electrical Engineering, Korea Advanced Institute of Science and Technology (KAIST), Daejeon, South Korea, in 2015, 2017, and 2021, respectively. 

From 2015 to 2017, he was also with Healthrian R\&D Center, Daejeon, South Korea, where he designed bio-potential readout IC for mobile healthcare solutions. In 2020, he was a Visiting Student with the Institute of Neuroinformatics, University of Zurich and ETH Zurich, Zurich, Switzerland, where he is currently working as a Postdoctoral Researcher, since 2021. His research interests include analog/mixed-signal ICs for time-domain processing, in-memory computing, bio-impedance sensor, and neuromorphic audio sensor.
\end{IEEEbiography}

\begin{IEEEbiography}
[{\includegraphics[width=1in,height=1.25in,clip,keepaspectratio]
{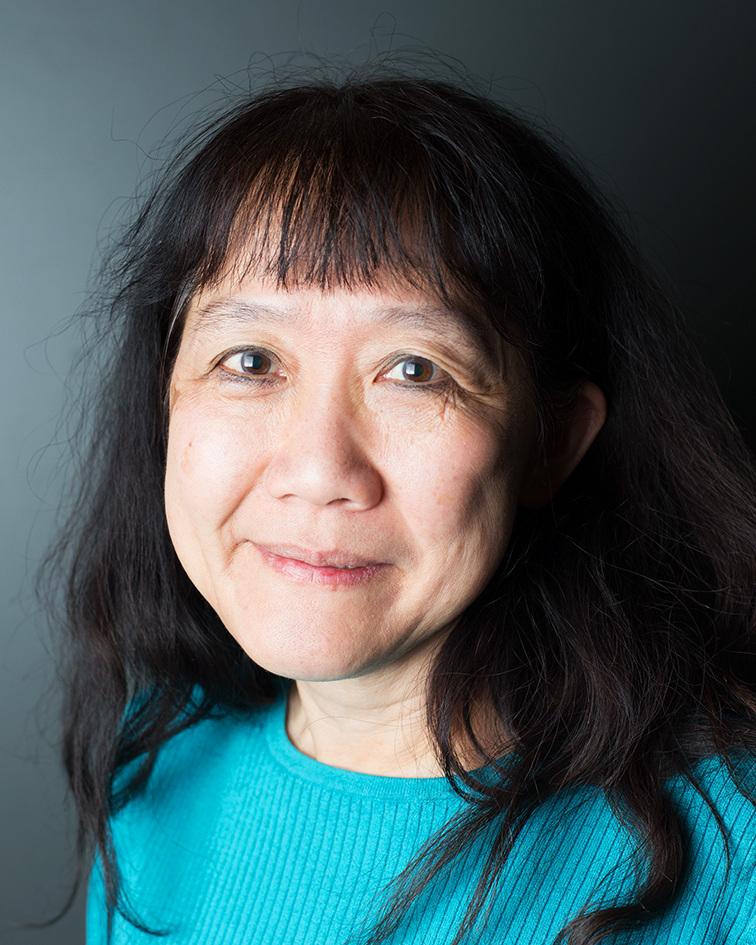}}]
{Shih-Chii Liu}
(Fellow, IEEE) received the bachelor’s degree in electrical engineering from the Massachusetts Institute of Technology, Cambridge, MA, USA, and the Ph.D. degree in the Computation and Neural Systems program from the California Institute of Technology, Pasadena, CA, USA, in 1997. She is currently a Professor in the Faculty of Science at the University of Zurich. She co-directs the Sensors group at the Institute of Neuroinformatics, University of Zurich and ETH Zurich. Her group’s research focuses on sensor integrated circuit designs including the spiking silicon cochlea and bio-inspired auditory sensors; and real-time energy-efficient hardware systems that combine both sensor and event-driven low-compute deep neural network algorithms, targeting always-on edge AI and wearable applications.
\end{IEEEbiography}

\end{document}